\documentclass[twocolumn]{aastex61}
\usepackage{booktabs}
\usepackage{hyperref}
\hypersetup{
    colorlinks=true,
    linkcolor=blue,
    filecolor=magenta,      
    urlcolor=cyan,
}
\usepackage{rotating}
\usepackage{multirow}
\def\arcsec{\ifmmode^{\prime\prime}\;\else$^{\prime\prime}\;$\fi}
\def\arcmin{\ifmmode^{\prime}}

\received{??} 
\revised{??}
\accepted{??}
\submitjournal{ApJ}

\shorttitle{AGN torus at the sub-millimeter}
\shortauthors{A. Pasetto}

\begin{document}
\title{AGN torus detectability at sub-millimeter wavelengths: \\ What to expect from ALMA continuum data}

\correspondingauthor{A. Pasetto, C\'atedra CONACyT}
\email{a.pasetto@irya.unam.mx}

\author{Alice Pasetto}
\author{Omaira Gonz\'alez-Mart\'in}
\author{Donaji Esparza-Arredondo}
\author{Natalia Osorio-Clavijo}
\author{Cesar Victoria-Ceballos}

\affil{Instituto de Radioastronom\'ia y Astrof\'isica (IRyA-UNAM), 3-72 (Xangari), 8701, Morelia, Mexico}

\author{Mariela Mart\'inez-Paredes}

\affil{Instituto de Radioastronom\'ia y Astrof\'isica (IRyA-UNAM), 3-72 (Xangari), 8701, Morelia, Mexico}
\affil{Korea Astronomy and Space Science Institute 776, Daedeokdae-ro, Yuseong-gu, Daejeon, Republic of Korea (34055)}
\begin{abstract}
Dust close ($\sim$few pc) to the accretion disk in active galactic nuclei (AGN) is key to understand many of their observational signatures and it is key to trace how the AGN is fed or even evolve along its duty cycle. With estimated sizes of less than 10 pc, as constrained by mid-infrared high angular resolution data, only the superb spatial resolution achieved by ALMA is able to actually image this dusty structure. However, the question regarding the conditions in which the dust at sub-milimeter (sub-mm) wavelengths where ALMA operates, arises.   We study the detectability of the emission associated with the AGN dusty structure at sub-mm wavelengths using ALMA, in a theoretical and observational way. Theoretically, we use the Clumpy models from Nenkova et al. together with the mid-infrared to X-ray and the radio fundamental plane scaling relations. We find that it is more likely to detect bigger and denser dusty tori at the highest ALMA frequency (666 GHz/450 $\mu$m). We also find that with 1h at 353 GHz/850 $\mu$m and 10h at 666 GHz/450 $\mu$m we can detect, with a high detection limit, a 1 mJy torus (characteristic of bright AGN). This means, an object for which the unresolved SED at 12 $\mu$m has a flux $\sim1$ mJy.
Observationally, we use four prototypical AGN: NGC\,1052 (low-luminosity AGN), NGC\,1068 (Type-2), NGC\,3516 (Type 1.5), and IZw1 (QSO), with radio, sub-millimeter, and mid-IR data available. All the mid-infrared spectra are best fit with the smooth model reported by Fritz et al. A power law and a single, or a composition of, synchrotron component/s reproduce the cm radio wavelengths. We combined and extrapolated both fits to compare the extrapolation of both torus and jet contributors at sub-mm wavelengths with data at these wavelengths. Our observational results are consistent with our theoretical results. The most promising candidate to detect the torus is the QSO IZw1 (therefore, highly accreting sources in general), although it cannot be resolved due to the distance of this source. We suggest that to explore the detection of a torus at sub-mm wavelengths, it is necessary to perform an SED analysis including radio data, with particular attention to the angular resolution.
\end{abstract}

\section{Introduction}\label{sec:intro}

Although the unification scheme \citep[e.g.,][]{Antonucci93, Urry_Padovani95} of active galactic nuclei (AGN) may not be universal, the most accepted description is that dust surrounding the central engine is key to explain some observational differences between AGN classes. As a first approach, the point of view of the observer with respect to this dusty structure is able to explain the type-1/type-2 dichotomy \citep{Pier93, Granato_Danese94, Efstathiou_Rowan-Robinson95}. 

The morphology and properties of this dusty structure are far from being understood \citep[e.g., see][]{Netzer15}. The torus dust grains absorb ultraviolet photons from the central engine and re-radiate them in the infrared (IR), peaking at mid-IR wavelengths around 5-35 $\rm{\mu m}$ \citep[e.g.,][]{Sanders89,Sargsyan11, Shi14}. Thus, from the observational point of view, many studies have been devoted to study the Spectral Energy Distribution (SED) of the dusty region at near- and mid-IR wavelengths. Several model fittings aimed to extract physical and geometrical properties from unresolved SED, by using high angular resolution data (subarcseconds) \citep[e.g.,][]{Ramos-Almeida09, Ramos-Almeida11,Alonso-Herrero11, Martinez-Paredes15, Fuller16, Lopez-Rodriguez16, Martinez-Paredes17, Gonzalez-Martin17, Garcia-Burillo17} and, in some cases, milliarcseconds interferometric observations \citep[e.g.,][]{Hoenig10A, Tristram11}. They found that for Seyferts the dusty structure should be concentrated into the inner $\rm{<}$5\,pc, as derived from near- and mid-IR wavelengths \citep{Radomski08, Ramos-Almeida09,Ramos-Almeida11,Alonso-Herrero11}, that extend for the highest AGN luminosities \citep{Martinez-Paredes17} and it is even smaller than 1\,pc for low-luminosity AGN \citep[LLAGN,][]{Gonzalez-Martin17}. 


\citet{Ramos-Almeida14} found that a minimal set of data (J+K+M band photometry + N-band spectroscopy) is necessary to constrain the geometrical parameters of the dusty torus using clumpy models for Seyfert galaxies. However, to have a reliable estimate of the torus size at mid-infrared, it is necessary to add the FIR emission to the SED in order to probe the coolest dust of the torus \citep{Garcia-Burillo16, Fuller16, Lopez-Rodriguez18}. This implies the need of FIR sub-arcsec resolution data. Infrared interferometry constitutes one of the most precise ways of characterizing the dusty structure using observations by comparing them with torus models \citep[][]{Hoenig06,Tristram11}. However, the signal-to-noise ratio (S/N) required for, e.g., the Very Large Telescope Interferometer (ESO-VLTI) observations strongly restricts the number of observable sources.

This has changed with the new sub-millimeter (sub-mm) capabilities as Atacama Large Millimeter Array (ALMA) or Large Millimeter Telescope (LMT). High resolution sub-mm observations are important in probing the morphology, column density and dynamics of obscuring matter in AGN. ALMA is able to achieve 0.02 arcsec resolution continuum images at $\rm{>400 \mu m}$. 
Very recently, some works attempted to detect and study the dusty torus, e.g., \cite{Izumi2018} studied the dusty central region of the Circinus galaxy testing the muti-phase dynamic torus model and \cite{Combes2018} explored the close environment around 7 seyfert/LINER galaxy detecting a molecular torus. However, none of them achieved to detect the inner dusty structure. Only \citet{Garcia-Burillo16} claimed the detection of an elongated continuum structure of $\rm{4\times7\,pc^2}$ associated to the dusty torus \citep[see also][]{Gallimore16}.

However, two main issues need to be faced when using sub-mm telescopes to study the dust in AGN: (1) Sensitivity issue: the peak of the dusty emission is below 30 $\rm{\mu m}$. The total flux expected above $\rm{400 \mu m}$ must be only a small fraction of what it is observed at mid-infrared wavelengths. Thus, although ALMA has a very good sensitivity limit, a proper study on the detectability of the dusty continuum at sub-mm wavelengths needs to be addressed. \citet{Aalto17} were not able to detect the dusty structure in the radio-quiet NGC\,1377 using a similar waveband compared to that used by \citet{Garcia-Burillo17}; and (2) Jet contribution: most of these AGN show a strong synchrotron contribution coming from the radio jet. This contribution is expected to be dominant at radio frequencies but a tail of this emission can contribute at other wavelengths, with non-negligible contribution to the sub-mm wavelengths. \citet{Izumi17} studied the LLAGN NGC\,1097 with less than 10 pc resolution at $\rm{\sim 800 \mu m}$. Indeed, they claimed that most of the sub-mm continuum emission is dominated by the jet thanks to the study of long-term variations at sub-mm wavelengths. 

It is the aim of this paper to study the detectability of extended emission associated to AGN dusty structure at sub-mm wavelengths using ALMA. For this purpose we will consider the most used SED models for dusty AGN emission as the Clumpy models \citep{Nenkova08A,Nenkova08B} together with known scaling relations between the accretion disk luminosity and the torus continuum \citep[][]{Gandhi09,Netzer15,Asmus15}, and the accretion disk luminosity, jet contribution, and super massive black-hole (SMBH) mass \citep[][]{Bonchi13,Saikia15} (see Section \ref{sec:theory}). 
In section \ref{sec:data}, we study the sensitivity and contributors to the sub-mm wavelengths of four prototypical AGN, covering a wide range of AGN classes: NGC\,1052 (LLAGN), NGC\,1068 (Type-2 Seyfert), NGC\,3516 (Type-1.5 Seyfert) and  I\,Zw1 (QSO). Details on the target selection are given in Section \ref{sec:casesofstudy}.
Using these targets, we will study the detectability of the sub-mm torus with available archival and/or literature data, combining together mid-IR, radio and sub-mm continuum flux densities (see Section \ref{sec:data}).
Our findings are discussed in Section \ref{sec:discussion} and a short summary of results is given in Section \ref{sec:summary}. We use the standard cosmology with $\rm{H_{0}=70  km/s/Mpc}$ along the text.



\begin{figure*}
\begin{center}
\includegraphics[width=2.0\columnwidth]{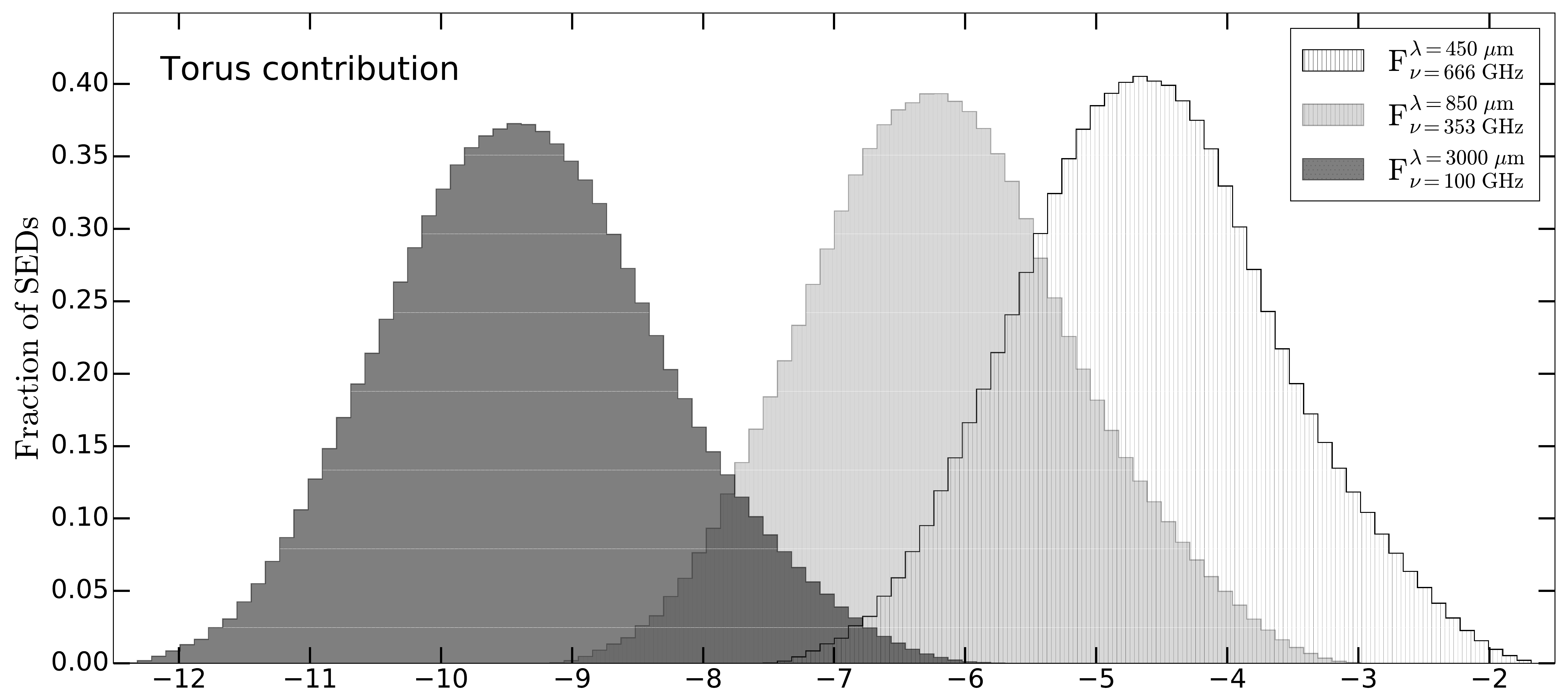}
\includegraphics[width=2.0\columnwidth]{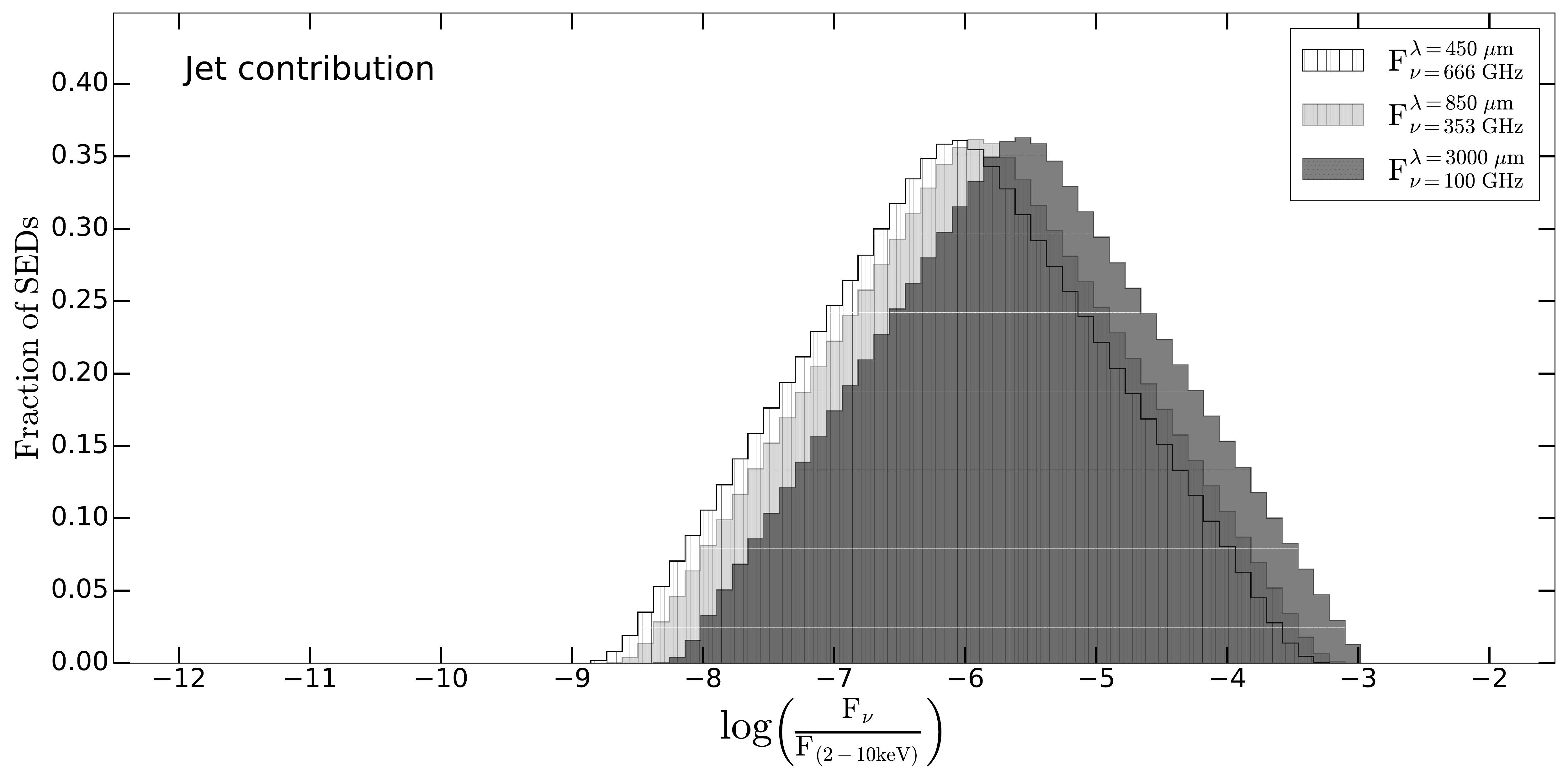}
\caption{Torus (top) and jet (bottom) flux contributions to the sub-mm wavelengths compared to the X-ray luminosity. The torus contribution is extrapolated from the Clumpy torus libraries, using the X-ray to mid-IR relation. The jet contribution is computed assuming a single synchrotron component and the radio fundamental plane (see text). }
\label{fig:SEDcontributors}
\end{center}
\end{figure*}

\begin{figure*}
\begin{center}
\includegraphics[width=0.75\columnwidth]{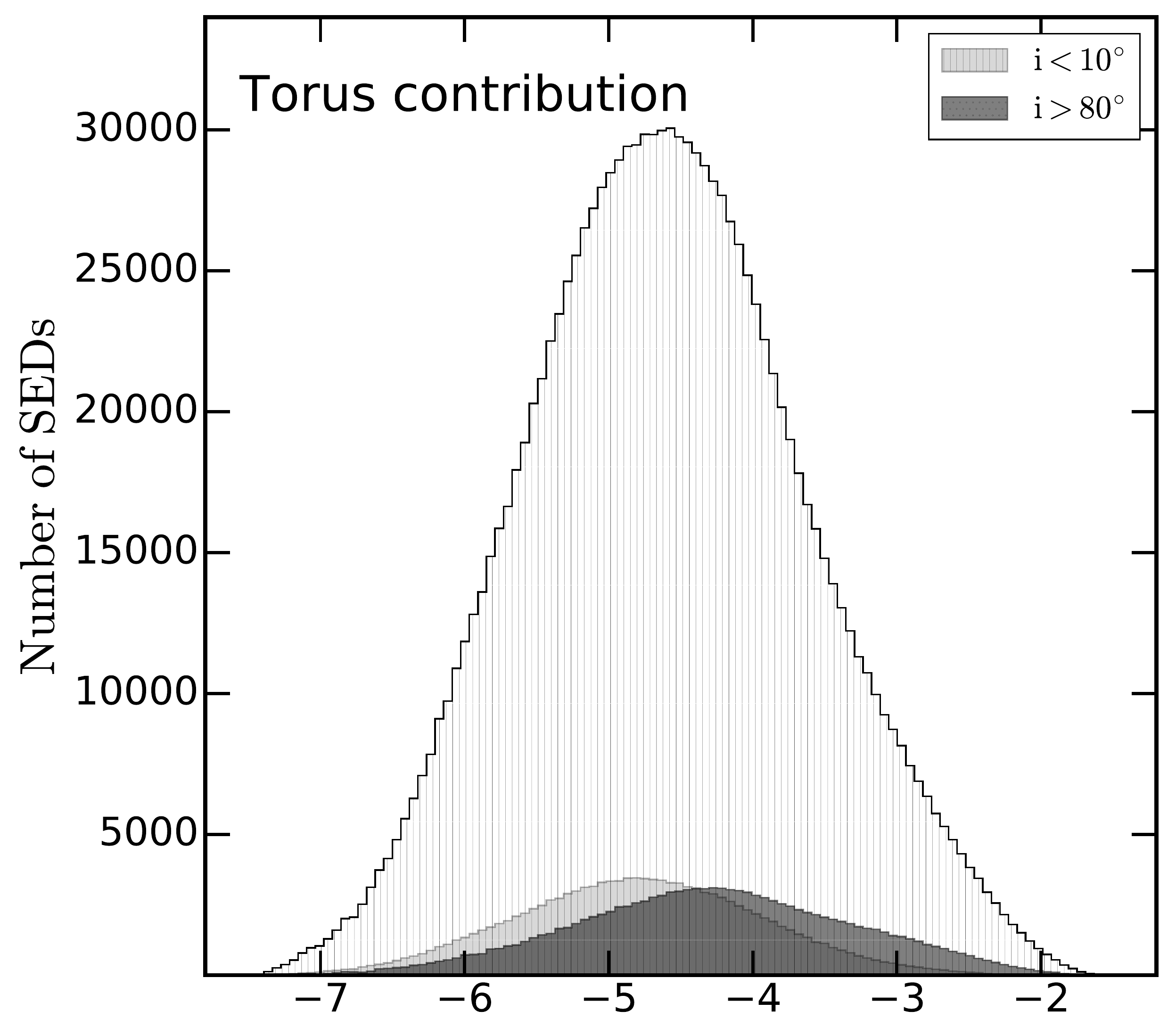}
\includegraphics[width=0.62\columnwidth]{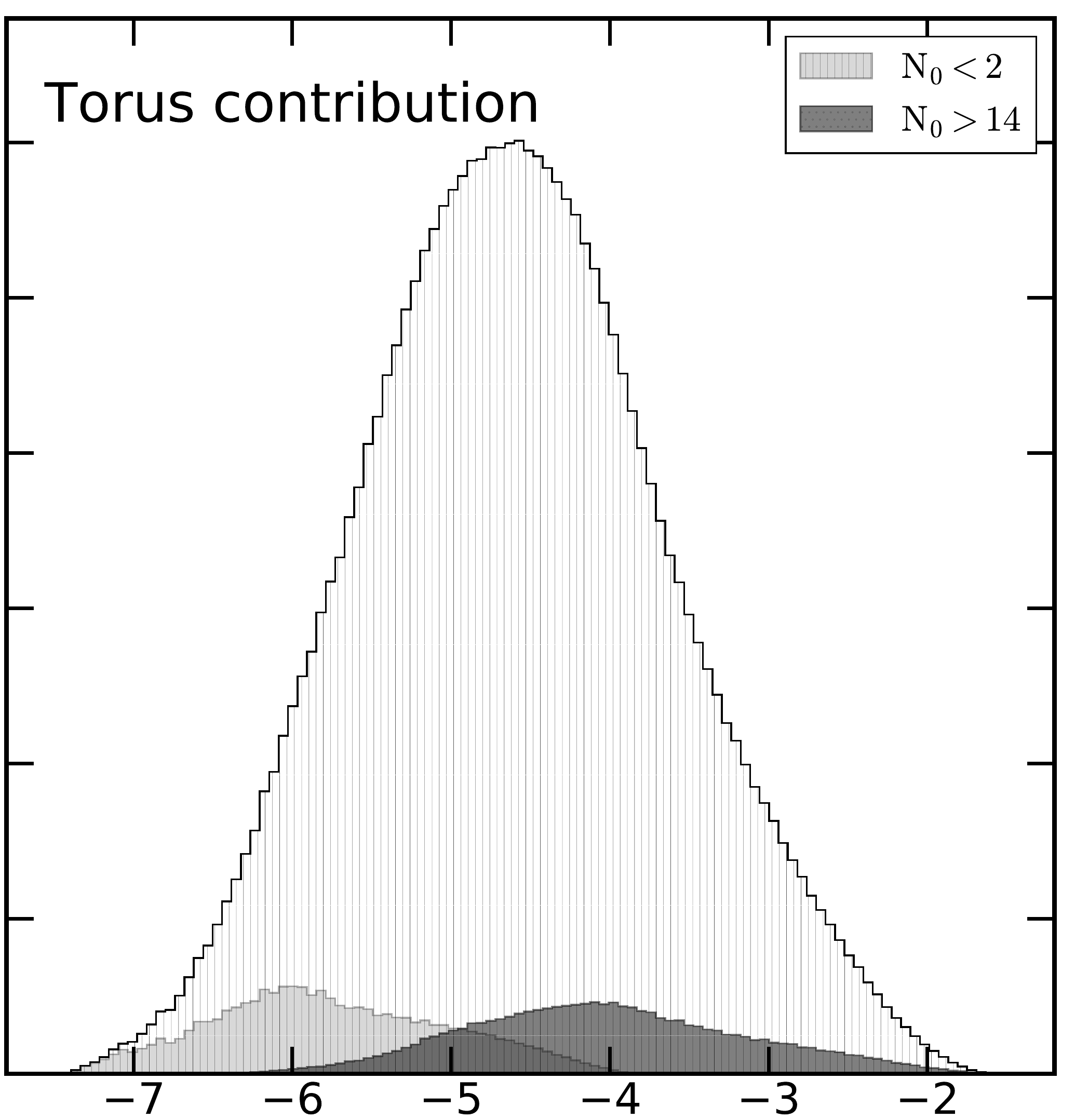}
\includegraphics[width=0.62\columnwidth]{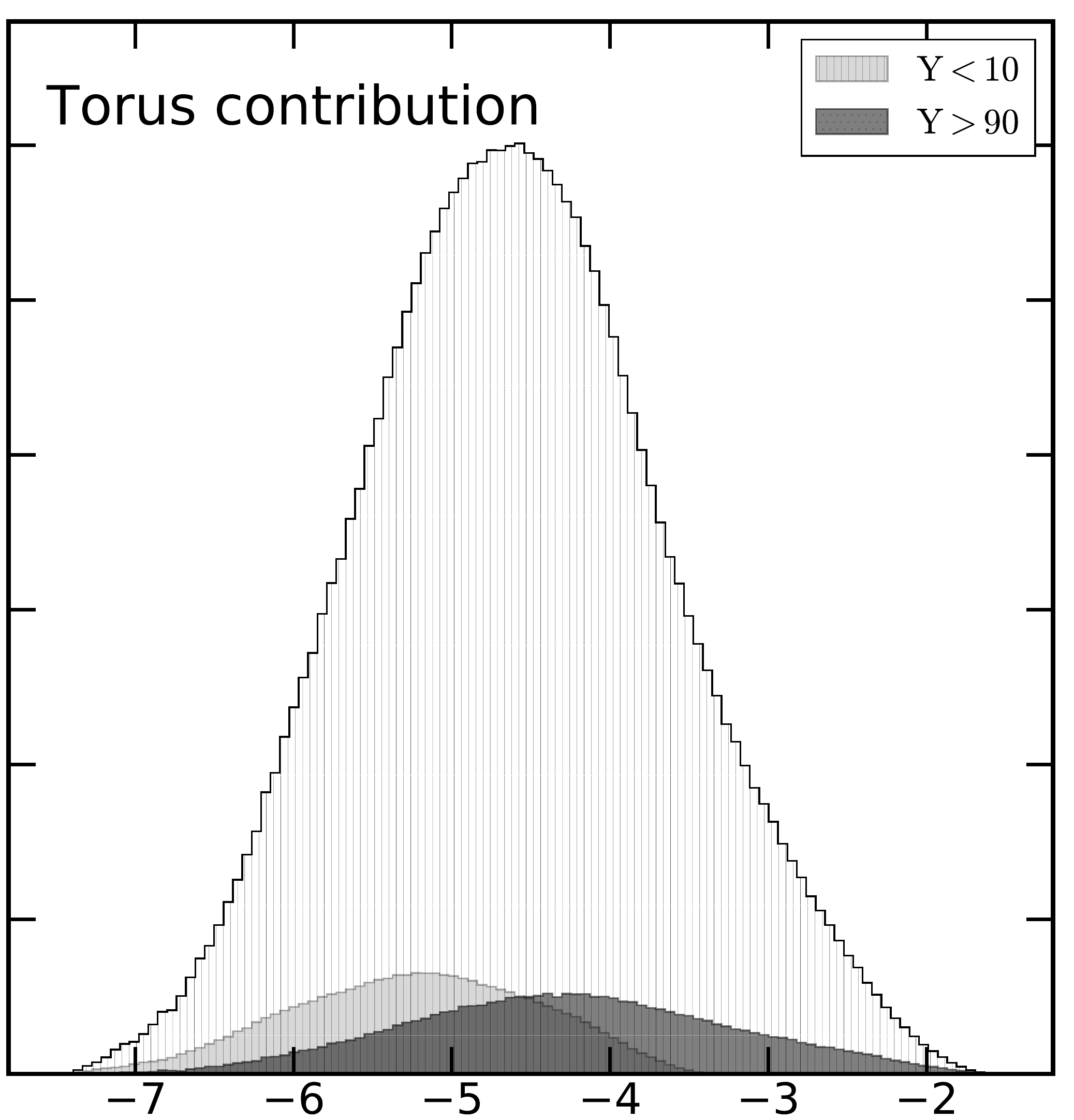}
\includegraphics[width=0.75\columnwidth]{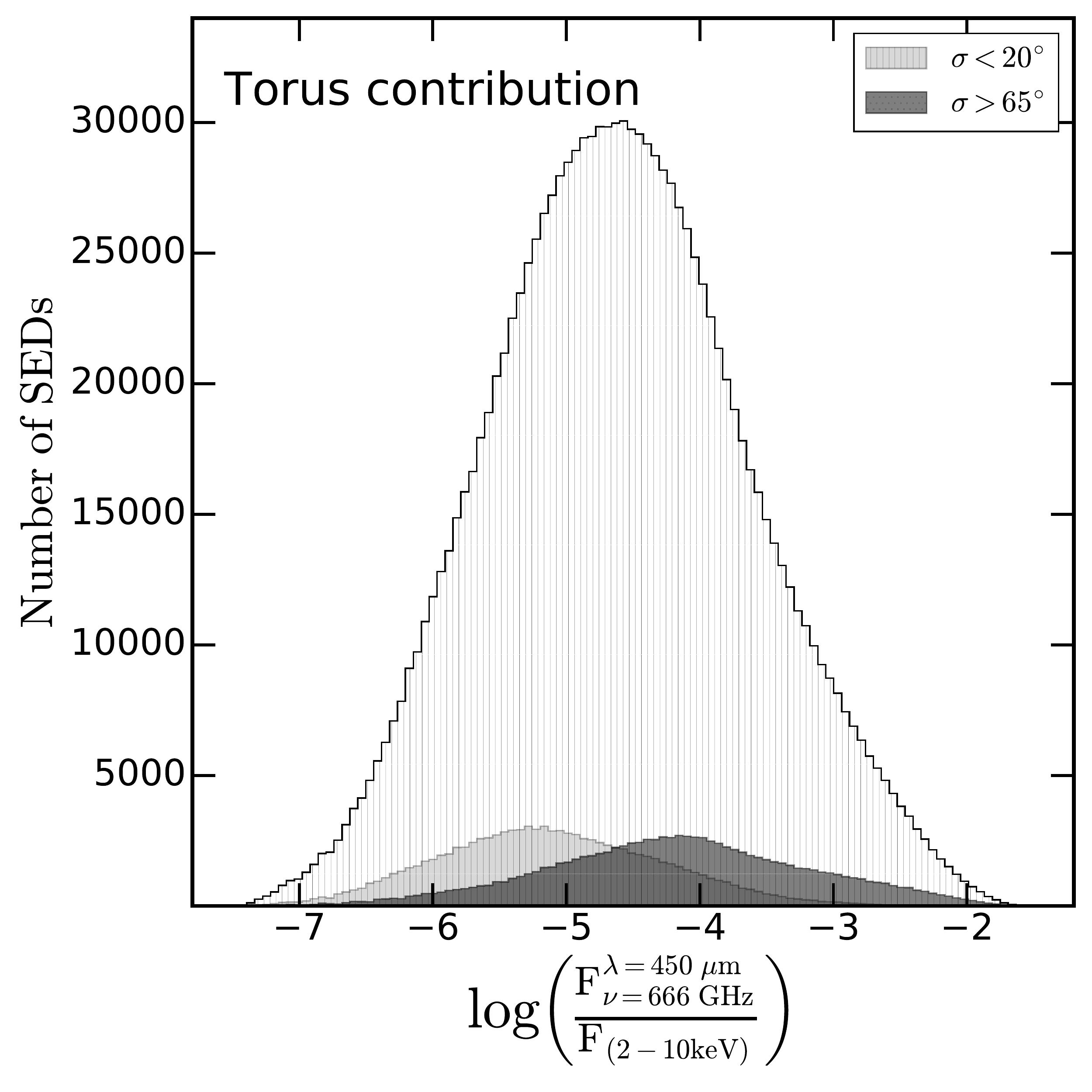}
\includegraphics[width=0.62\columnwidth]{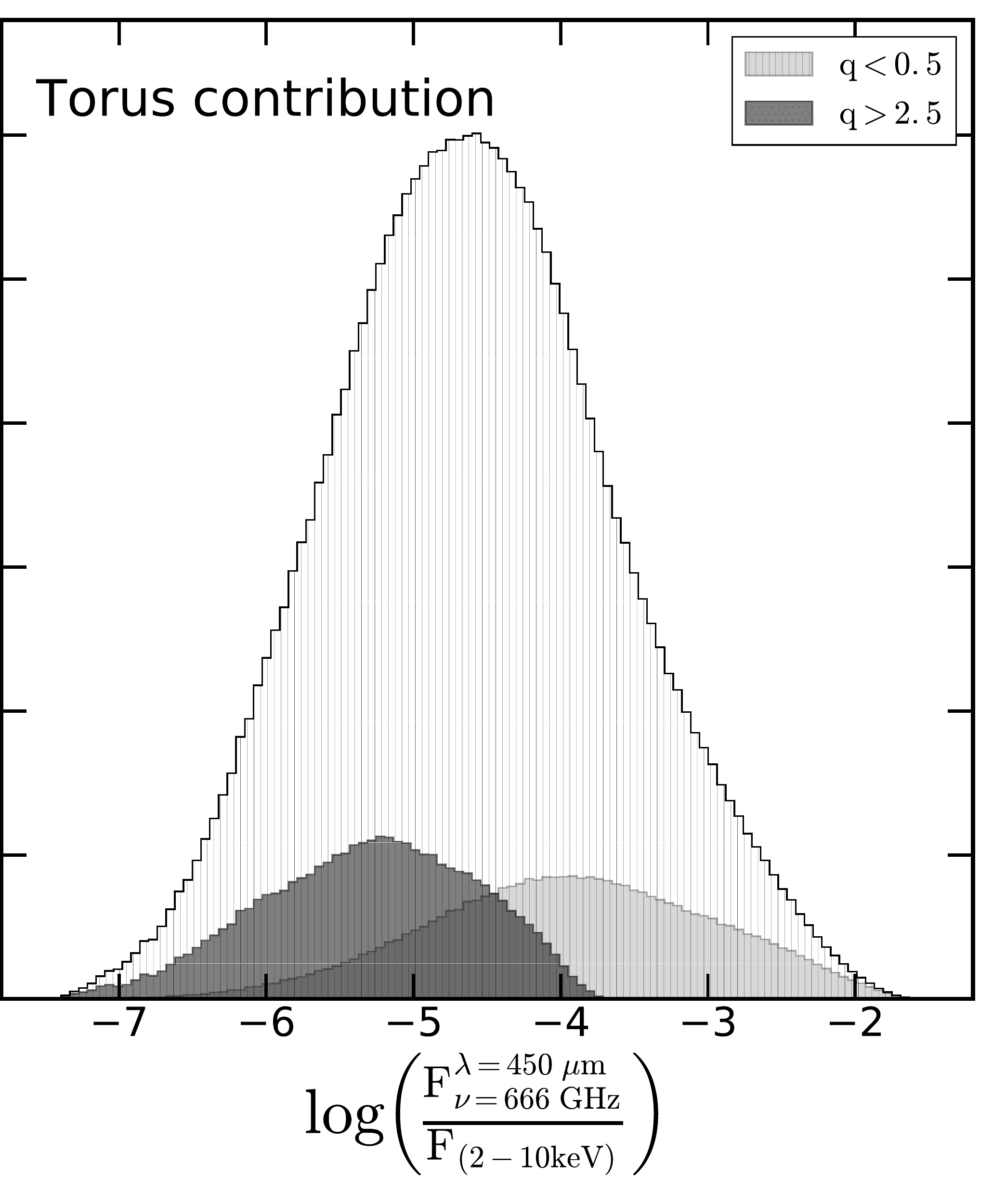}
\includegraphics[width=0.62\columnwidth]{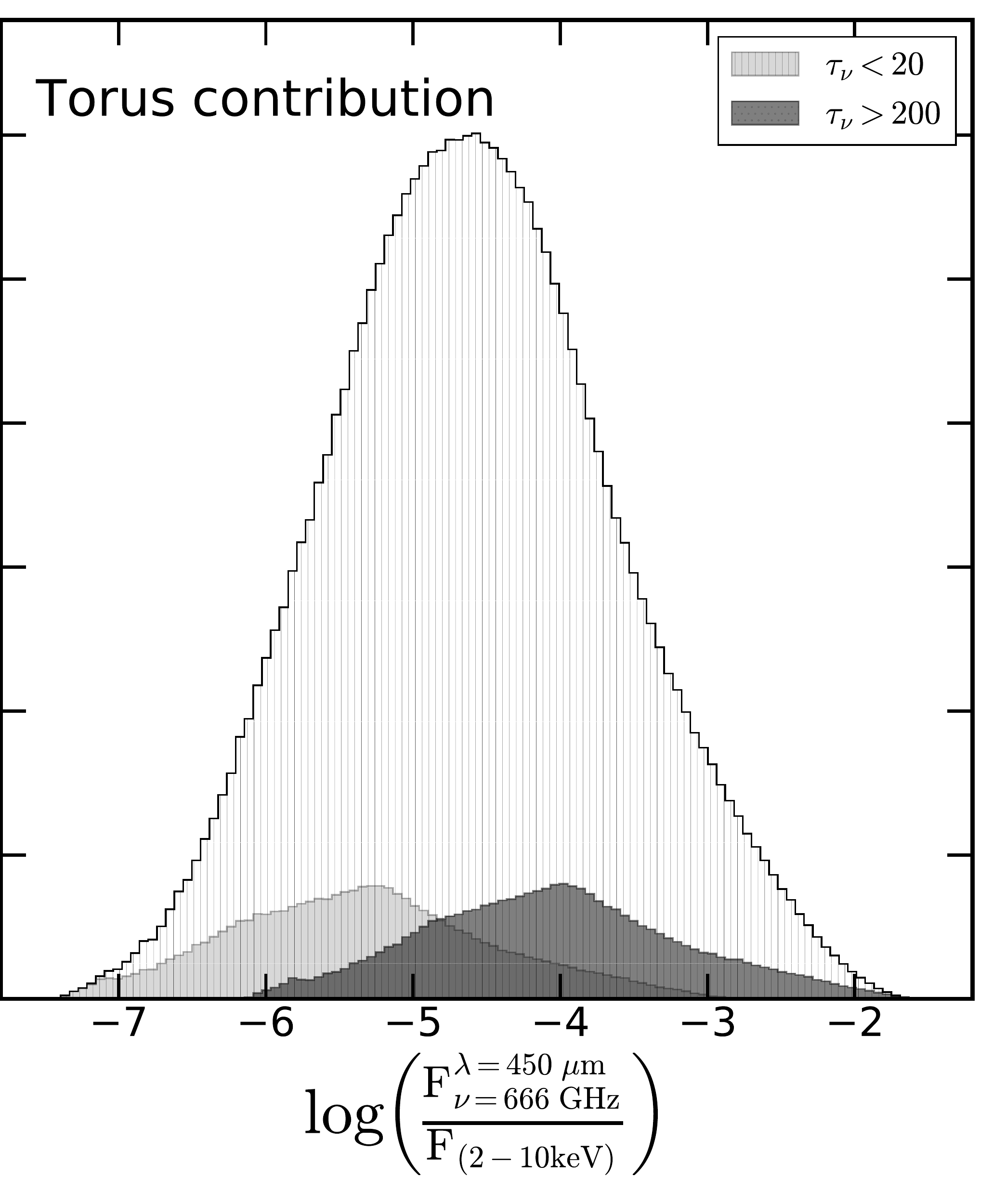}
\caption{Distribution of the torus contributions at 450$\rm{\mu m}$ compared to the 2-10 keV X-ray flux density depending on the inclination angle \textit{i} (top-left panel), equatorial number of clouds \textit{N$_{0}$} (top-middle panel), torus outer radius \textit{Y} (top-right panel), half openning angle of the torus \textit{$\sigma$} (bottom-left panel), slope of the radial distribution of clouds \textit{q} (bottom-middle panel), and the optical depth of the clouds \textit{$\tau_{\nu}$} (bottom-right panel).}
\label{fig:Toruscontributors}
\end{center}
\end{figure*}

\begin{figure*}
\begin{center}
\includegraphics[width=1.0\columnwidth]{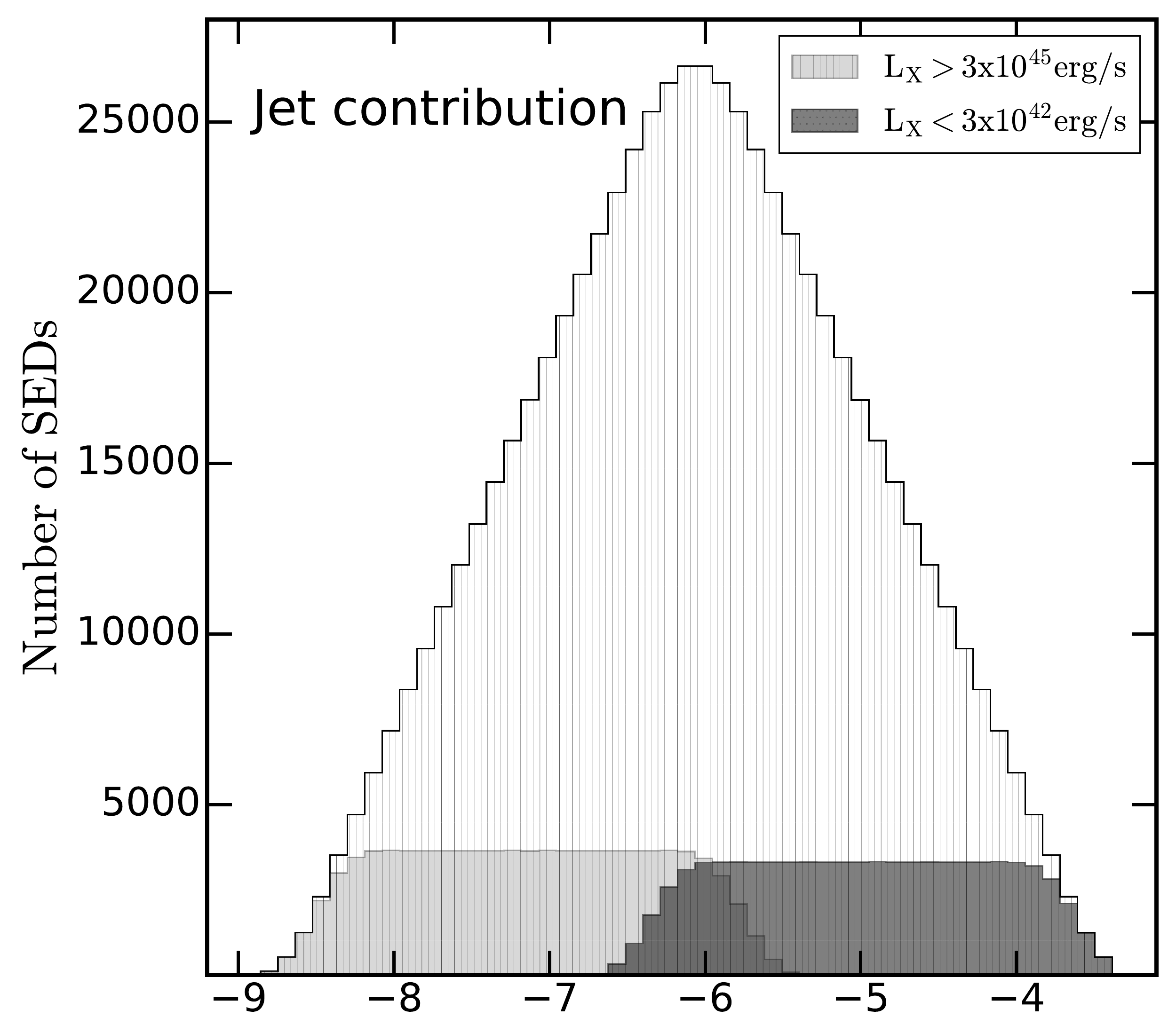}
\includegraphics[width=0.81\columnwidth]{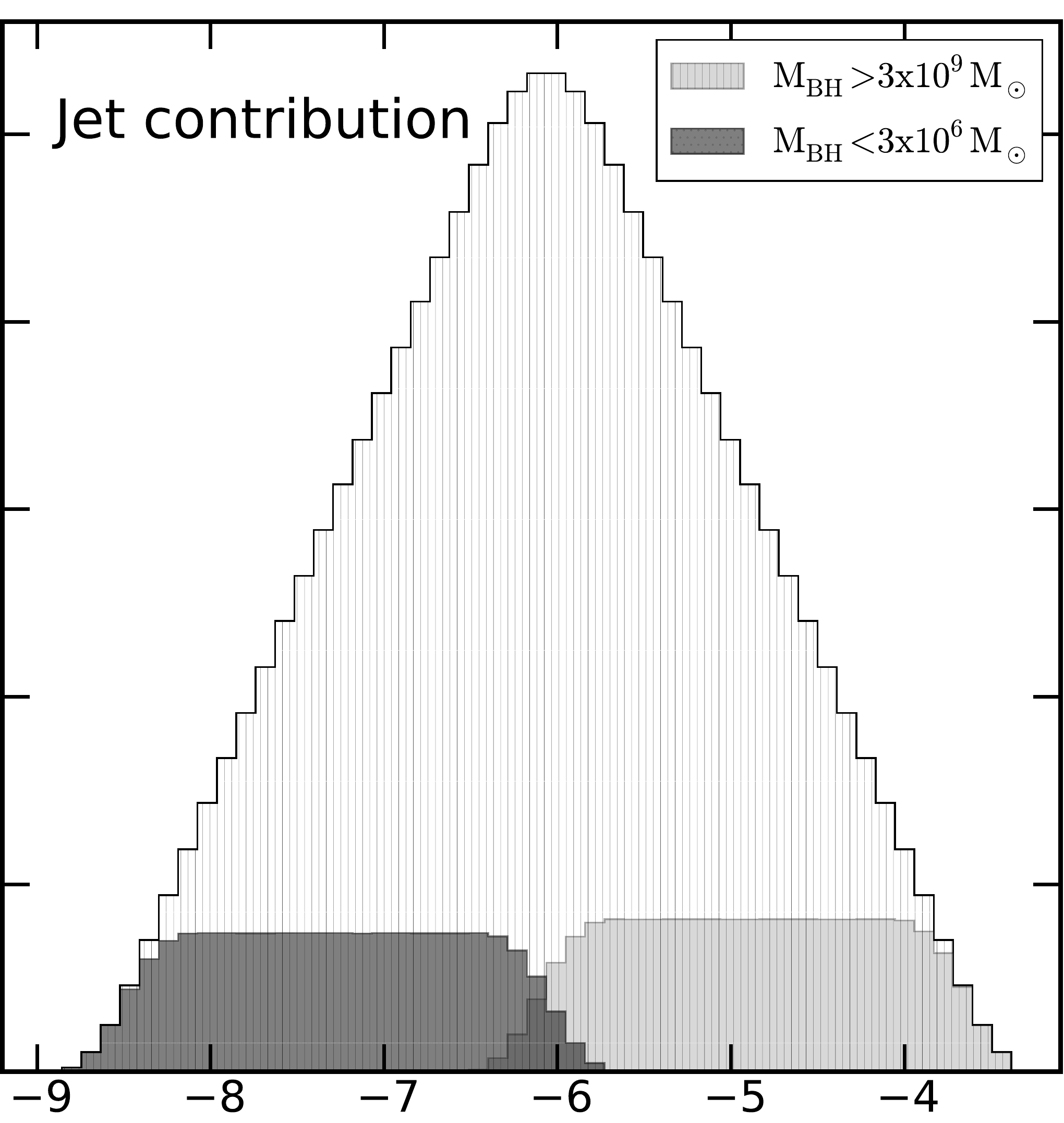}
\includegraphics[width=1.0\columnwidth]{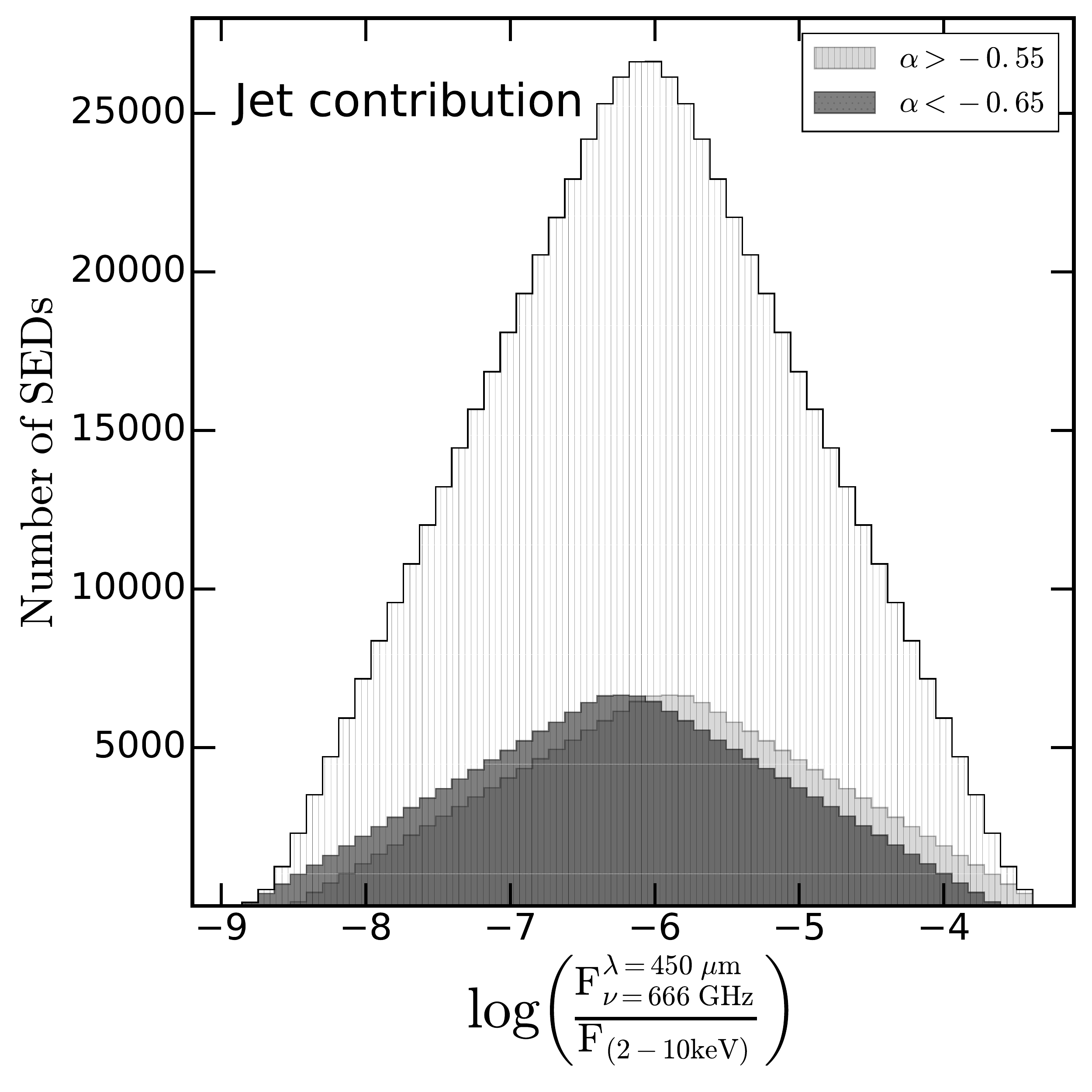}
\includegraphics[width=0.81\columnwidth]{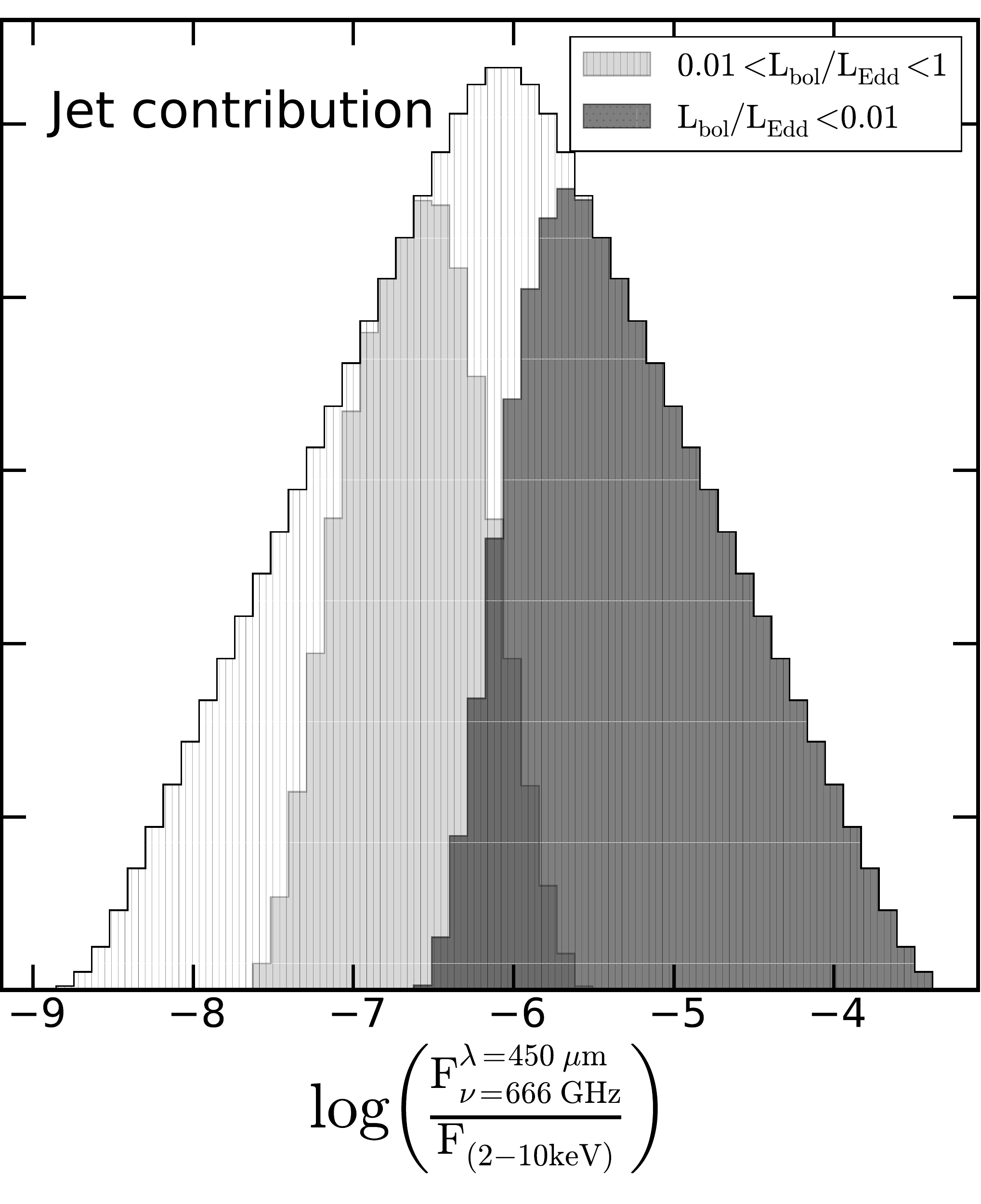}
\caption{Jet contributions at 666GHz/450$\rm{\mu m}$ depending on the X-ray luminosity (top-left panel), BH mass (top-right panel), the synchrotron spectral index (bottom-left panel), and the accretion rate(bottom-right panel).}
\label{fig:Jetcontributors}
\end{center}
\end{figure*}

\section{Torus detection with ALMA through theoretical predictions}\label{sec:theory}

We compute the expected contributions to the sub-mm range as the ratio between the flux at sub-mm wavelengths and that at X-ray wavelengths. This allows us to scale the sub-mm flux density to the accretion disk flux density. We use the integrated luminosity in the 2-10 keV band (hereinafter $\rm{L_{[2-10keV]}}$) for the X-ray luminosity and the monochromatic luminosities at 100\,GHz (3000\,$\rm{\mu m}$, ALMA band 3), 353\,GHz (850\,$\rm{\mu m}$, ALMA band 7), and 666\,GHz (450\,$\rm{\mu m}$, ALMA band 9) for the sub-mm wavelengths. Hereinafter we refer to the fluxes at these bands as $\rm{F_{\nu=100 GHz}^{\lambda = 3000 \mu m}}$, $\rm{F_{\nu=353 GHz}^{\lambda = 850 \mu m}}$, $\rm{F_{\nu=666 GHz}^{\lambda = 450 \mu m}}$, respectively. We chose these bands because they cover the full waveband covered by ALMA. These three bands can achieve an spatial resolution of 0.2, 0.06, and 0.03\arcsec, respectively, considering the largest configuration (C43-7 available during cycle 6). The 666\,GHz continuum emission (band 9) was chosen by \citet{Garcia-Burillo16} \citep[see also][]{Imanishi16a} and \citet{Aalto17} to identify the torus contribution to the continuum at sub-mm wavelengths for NGC\,1068 and NGC\,1377, respectively (see Section \ref{sec:intro}). In the same manner, \citet{Izumi17} used 350\,GHz (band 7) to try to recover the torus contribution to the sub-mm wavelengths for NGC\,1097. We also test band 3 (100\,GHz) because it provides slightly better sensitivity compared to the other bands. 

Note that in this section we do not take into account the dilution, although it
might be important if spatial resolution is moderate (see Section
\ref{sec:data}, e.g. for the case of NGC\,1068). For instance, the MIR flux will change if the field of
view of the observations change \citep[e.g][]{Tristram14, Prieto10}

	\subsection{Torus contribution}\label{sec:toruscontrib}

In order to estimate the torus contribution to the sub-mm range we estimate two ratios: (1) the 12\,$\rm{\mu m}$ to 2-10 keV flux ratio ($\rm{F_{12\mu m}/F_{[2-10 keV]}}$) and (2) the sub-mm to the 12\,$\rm{\mu m}$ flux ratio ($\rm{F_{\nu}^{\lambda}/F_{12\mu m}}$). 

We use the well established mid-IR to X-ray scaling relation for AGN \citep{Lutz04, Gandhi09} to estimate the $\rm{F_{12\mu m}/F_{[2-10 keV]}}$ ratio. The most updated version of this relation is presented by \citet{Asmus15} using a sample of 152 nearby AGN, and can be written as: 

\begin{eqnarray}
log \left( \frac{L^{nuc}_{12 \mu m}}{10^{43} erg s^{-1}} \right) = (0.30\pm0.03) + \nonumber \\  (0.98\pm0.03) log \left( \frac{L^{int}_{[2-10] KeV}}{10^{43} erg s^{-1} }\right) 
\end{eqnarray}
We then compute the expected $\rm{F_{\nu}^{\lambda}/F_{12\mu m}}$ using the SEDs of clumpy torus model produced by \citet{Nenkova08B}. This clumpy torus model depends on six parameters: the observer inclination angle towards the torus $\rm{i}$, half angular width of the torus $\rm{\sigma}$, outer radius of the torus scaled to the inner radius $\rm{Y=R_{out}/R_{in}}$\footnote{The inner radius of the torus is set to the sublimation radius of the dust that depends on the accretion disk bolometric luminosity}, the number of clouds at the equatorial plane of the torus $\rm{N_{0}}$, the steepness of the radial distribution of clouds $\rm{q}$, and the opacity of the individual clouds $\rm{\tau_{\nu}}$. Thus, our $\rm{F_{\nu}^{\lambda}/F_{12\mu m}}$ ratio depends on these six parameters. Note that the SEDs of the clumpy torus model cover up to 1000$\rm{\mu m}$. We linearly extrapolate the SEDs above 1000$\rm{\mu m}$ to estimate the expected flux at 3000$\rm{\mu m}$ (i.e. band 3 at 100\,GHz). 

The distribution of torus $\rm{F_{\nu}^{\lambda}/F_{[2-10keV]}}$ ratios are shown in the top panel of Fig.\ \ref{fig:SEDcontributors}. As expected, the highest torus contribution is obtained in the 666\,GHz (450\,$\rm{\mu m}$), with a sub-mm flux density $\rm{\sim 10^{-6}-10^{-3}}$ times the X-ray flux density. 

\begin{figure}[ht!] 
	\begin{center}
		\includegraphics[width=1.0\columnwidth]{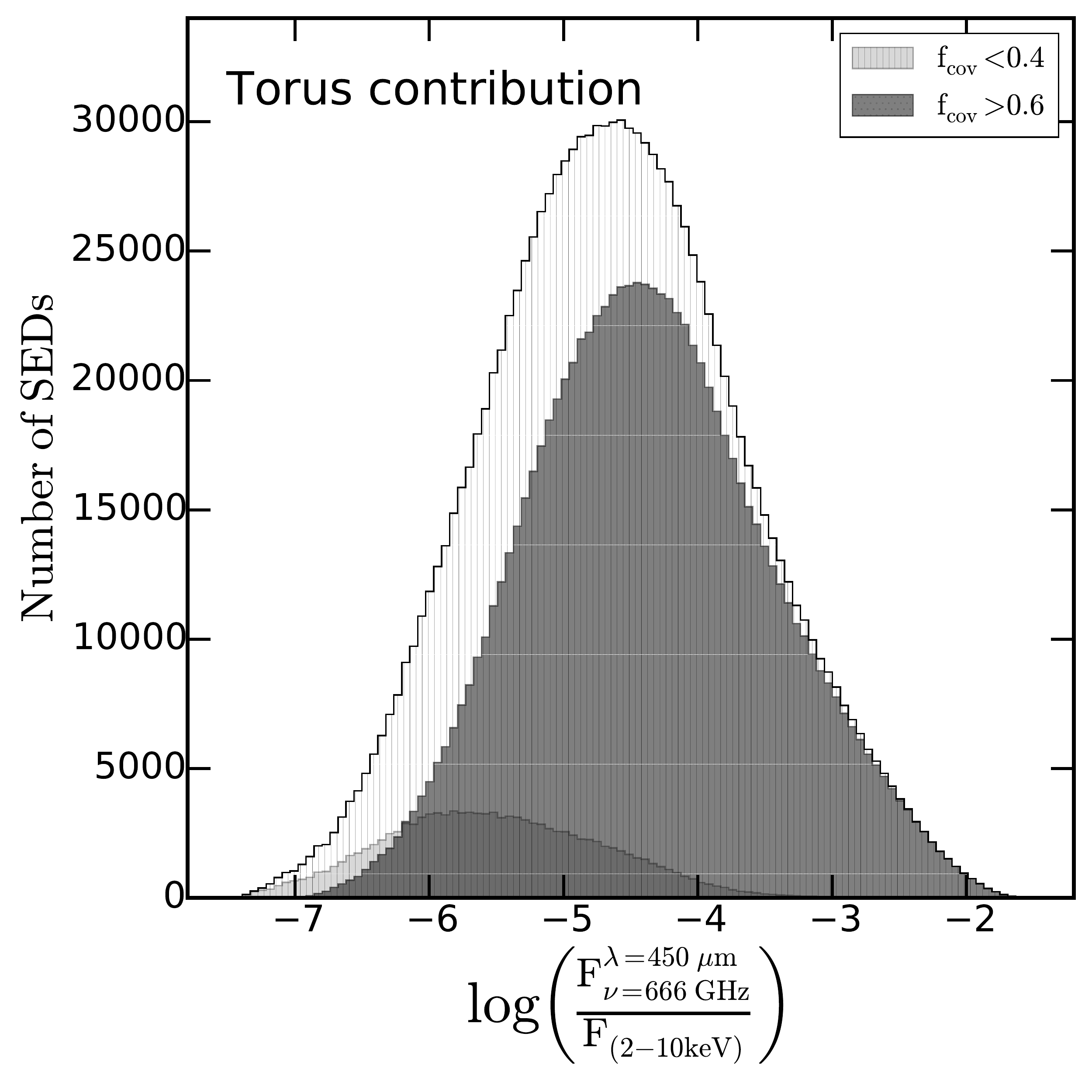}
\caption{Histogram of the covering factor distribution. Values larger than f$_{cov}$ $>$ 0.6
could be important for the contribution of the torus in the sub-mm band. However, due to its wide width,
this parameter is not determinant. Small values of the covering factor, f$_{cov}$ $<$ 0.4 are not determinant at all.  }
		\label{fig:histograms_torus_fcov}
	\end{center}
\end{figure}

We also show how the parameter space affects the torus contribution to the sub-mm wavelengths in Fig. \ref{fig:Toruscontributors}. Each panel shows the distribution of the $\rm{F_{\nu=666GHz}^{\lambda=450 \mu m}/F_{[2-10keV]}}$ ratio for the lowest (light gray) and highest (dark gray) values for each parameter. The highest values are obtained for  higher number of clouds in the equatorial plane ($\rm{N_0}$), larger outer radius of the torus compared to the inner one ($\rm{Y}$), larger half angular width of the torus ($\rm{\sigma}$), flatter radial distribution of clouds ($\rm{q}$), and larger values of the optical depth of individual clouds ($\rm{\tau_{\nu}}$). No significant effect is produced by edge-on or face-on torus (the inclination angle \textit{i}). All together means that, as expected, bigger and denser dusty tori are more likely to be detected at sub-mm wavelengths.
Finally we also considered how the  fraction of the sky obscured by dust, i.e. the covering factor, at the highest ALMA frequency can be incisive in the torus detectability and the result is shown in Fig.\ \ref{fig:histograms_torus_fcov}.  Large covering factor values (f$_{cov}$ $>$ 0.6) are those for which the torus contribution could be important in the sub-mm window. However, because of the wide width of the distribution it is worth notice that large covering factor is not a determinant condition.

\begin{figure}[ht!] 
	\begin{center}
		\includegraphics[width=1.0\columnwidth]{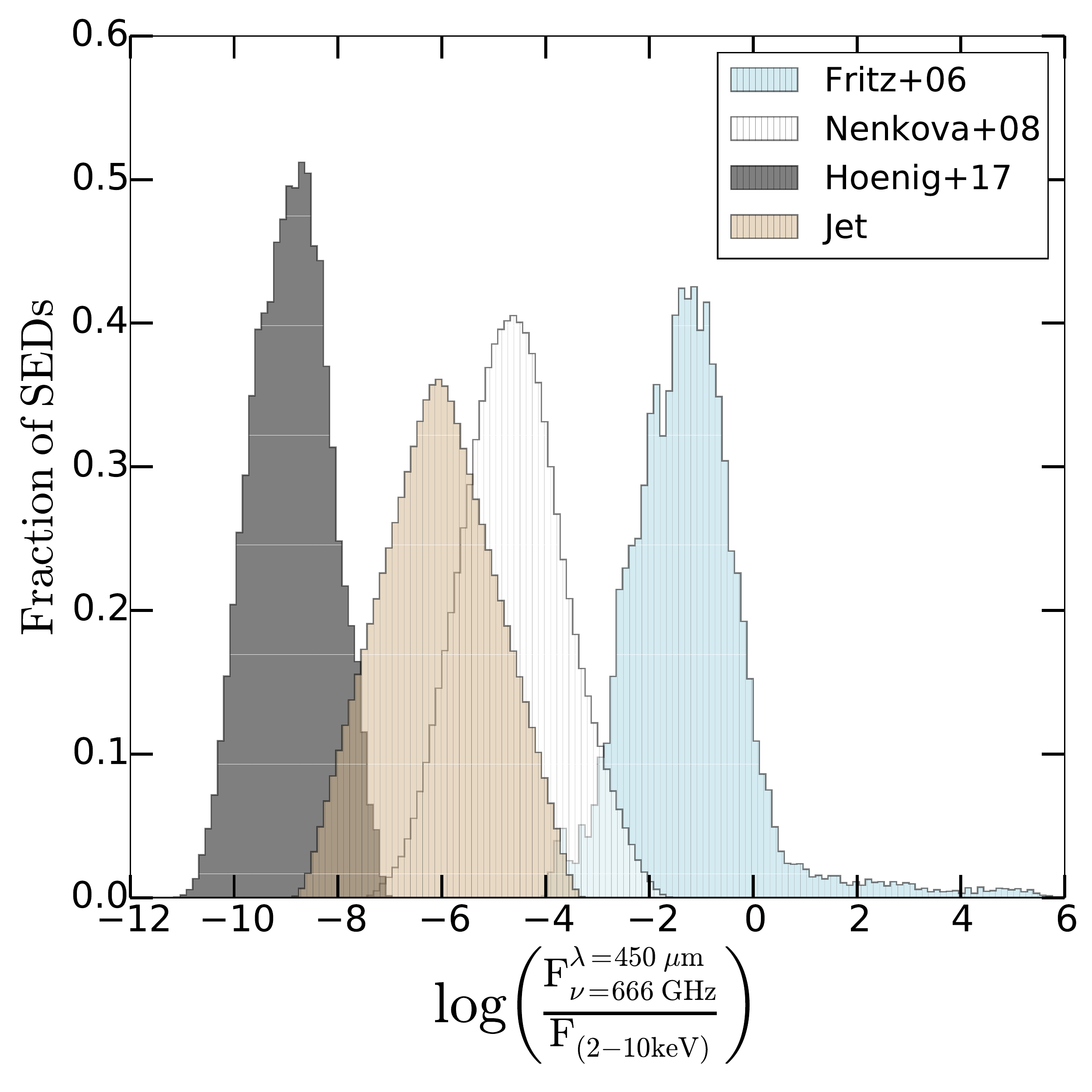}
\caption{Distribution of the disk+wind (gray), clumpy (white) and smooth (cyan) torus models together with the radio jet distribution (brown) at 666GHz/450$\mu$m. } 
		\label{fig:FullHistogramsTorusModel}
	\end{center}
\end{figure}

Finally, we investigate how these results could be influenced by the torus model selected. Fig.  \ref{fig:FullHistogramsTorusModel} shows the distribution of the clumpy torus
model by \cite{Nenkova08B}, the smooth torus model by \cite{Fritz06} and the disk + wind model by \cite{Hoenig17} together with the jet contribution at the highest ALMA frequency considered (666GHz/450$\mu$m, for which the highest torus contribution is obtained). The contribution of the AGN dusty torus depends on the distribution of the dust grains. For the case of a polar dusty wind torus distribution, the model predicts a negligible contribution of torus emission in the sub-mm band, while for dust grains distributed in a smooth torus , the model predicts larger contribution. Even if the most optimistic model, i.e. the smooth model, is used, the torus seems to be difficult to be detected at sub-mm wavelengths (see Sect. \ref{sec:data}). 
The histogram is also telling us that an important contribution from the synchrotron emission from the radio jet could contaminate the dust emission over the most common used clumpy torus.

Moreover, we also considered additional contribution from the hot
graphite dust and the dusty NLR clouds as proposed in several works to
describe the main emission components in the near and mid-IR spectra, e.g
\citep{Mor2009,Mor2012}.  The former contributes mainly a $\sim$ 3 $\mu$m
and it decays very steeply while the last could contribute to the mid-IR
spectrum. Indeed, it contributes above 10 micron and could be as large
as 40\% of the total emission at 24 $\mu$m.  Assuming this percentage of contribution remains the same at sub-mm wavelengths,  a shift of the torus models distributions in Fig. \ref{fig:FullHistogramsTorusModel} by a factor of 0.2 towards the right. 
Thus the NLR contribution do not have a large impact on the result presented above.

\subsection{Jet contribution}\label{sec:jetcontrib}
We extrapolate the contribution of the radio jet in the sub-mm range using the so-called fundamental plane relation \citep{Merloni2003}, for which radio luminosity (L$_{\nu}^{\lambda}$) is correlated with both the (hard) X-ray luminosity L$_{[2-10keV]}$ (from a corona closely linked to the accretion disk emission) and the mass of the SMBH (M$_{BH}$). This implies that a correlation between jet and disk flux density is unavoidable. \cite{Bonchi13} measured a fundamental plane for the largest  AGN-only sample for which radio, X-ray and K-band luminosity information exists. Moreover, they added radio upper limits and they considered a wide range of redshifts (up to z$\sim$ 5).
The fundamental plane relation obtained is the following:
\begin{eqnarray}
 \log L_{R} =  0.39\log(L_{[2-10keV]}) + 0.68 \log(M_{BH}) \nonumber \\ 
 + 16.61,
 \label{Bonchieq}
 \end{eqnarray}
\noindent
where L$_{R}$ is the 5-GHz nuclear luminosity in units of erg s$^{-1}$, with 1.4 GHz radio luminosities converted into 5 GHz luminosities assuming a radio spectral index $\alpha$ = --0.7 (where L$_{\nu}$ $\propto$ $\nu^{\alpha}$), L$_{[2-10keV]}$ is the 2-10 keV nuclear X-ray luminosity in units of erg s$^{-1}$, and M$_{BH}$ is the SMBH mass
in units of M $\rm{_{\odot}}$ \citep{Merloni2003}.
It is from the equation \ref{Bonchieq} that we extrapolate our expected sub-mm jet luminosity.  Therefore, we derive our sub-mm fundamental plane as:
\begin{eqnarray}
\log(L_{\nu}^{\lambda})= 0.39  \log(L_{[2-10keV]}) + 0.68  \log(M_{BH}) \nonumber \\ 
+ 16.61 - \alpha \log\left(\frac{5_{[GHz]}}{\nu}\right),
\end{eqnarray}
\noindent
where $\alpha$ represents the spectral index; i.e, the slope, for an optically thin radio component emitting synchrotron radiation, and $\nu$ are the three possible ALMA frequencies considered in this paper (i.e. 100 GHz, 353 GHz, and 666 GHz). Therefore, the ratio between F$_{\nu}^{\lambda}$/F$_{[2-10keV]}$ still depends on the X-ray luminosity L$_{[2-10keV]}$ as it follows:
\begin{eqnarray}
\log\left(\frac{F_{\nu}^{\lambda}}{F_{[2-10keV]}}\right)= C(\alpha, \nu) -0.61 \log(L_{[2-10keV]}) \nonumber \\ + 0.68 \log(M_{BH}) 
\end{eqnarray}
\noindent
where the constant C($\rm{\alpha}$, $\rm{\nu}$) = $\rm{16.61-\alpha log(5_{[GHz]}/\nu)}$.

For the purpose of our study we explore typical ranges for the parameters: spectral indexes for local AGN \citep{Condon2002,Sadler2014} of $\alpha$=[--0.5,--0.7], X-ray luminosity of L$\rm{_{[2-10keV]}}$=[10$^{42}$,10$^{46}$] erg s$^{-1}$, and SMBH masses of M$\rm{_{BH}}$=[10$\rm{^{6}}$, 10$\rm{^9}$] M$\rm{_{\odot}}$. The distribution of the sub-mm jet contribution peaks at around $\rm{\sim 10^{-6}}$, irrespective of the chosen sub-mm frequency (see bottom panel histogram in  Fig.\ \ref{fig:SEDcontributors}). Note that this value might be an overestimate of the jet contribution. \cite{Saikia2018}, investigated whether or not 1.4 GHz fluxes can trace nuclear activity \citep[using 10149 AGN taken from the Faint Images of the Radio Sky at Twenty-centimeters, FIRST, survey][]{White1997}. They concluded that the 1.4 GHz FIRST fluxes do not trace the pure "core" jet and instantaneous nuclear activity. \cite{Bonchi13}, mostly used data from the FIRST survey (with angular resolution of 5$\arcsec$) and they considered sources in a wide range of redshifts, meaning that several synchrotron components could contribute within the radio beam. Therefore, the radio flux density could be contaminated by radio extended emission, e.g., from the radio lobe, that can overestimate the radio flux density. This can produce an up-shift in the fundamental plane. In the practice, this means that pure core-jet component would have a lower flux density and the histograms showed in Fig.\ \ref{fig:SEDcontributors} (bottom panel) would move towards lower values. To isolate as much as possible the central radio components, observations at high resolution (e.g., using the JVLA in its A-configuration) and at high radio frequencies are needed.

We also show how the X-ray luminosity  L$\rm{_{[2-10keV]}}$, the SMBH
mass M$\rm{_{BH}}$, the synchrotron spectral index ($\alpha$), and the
accretion rate $\rm{L_{bol}/L_{Edd}}$ parameters affect the jet
contribution considering the highest frequency band (666GHz/450$\mu$m)
in Fig.\ \ref{fig:Jetcontributors}. The lowest jet contribution in this
band is obtained considering high X-ray luminosity, low SMBH masses,
and, therefore, high accretion rates. No significant effect is produced
by the narrow range of $\alpha$. It is worth to mention that a
spectral index of --0.5 or --0.7 are tracing similar class of objects.
\cite{Merloni2003} claimed that using a wide range of $\alpha$
(i.e., flat radio source with $\alpha$ $>$ --0.4 and steep radio sources
with $\alpha$ $<$ --0.4) the fundamental plane shows a considerable
scatter. In order to reduce this scatter they suggest to restrict to a
specific class of object, as we did considering the narrow $\alpha$
range of [--0.5, --0.7].
Also note that we exclude from this plot the lowest values of the total distribution because they are obtained with $\rm{L_{bol}}$ above the Eddington limit. Nevertheless, note that the radio jet contribution seems to be present for a wide range of cases in the sub-mm band, effectively showing $\rm{F_{\nu=666GHz}^{\lambda=450 \mu m}/F_{[2-10keV]}\sim 10^{-8}-10^{-4}}$.

\begin{table*}[t]
\scriptsize
\begin{center}
\begin{tabular}{l c c c c c c c c}
\hline \hline
Component		&		$\rm{\nu/\lambda}$				&	log($\rm{F_{\nu}^{\lambda}/F_{(2-10keV)}}$)	&	Percentage 	&	\multicolumn{2}{c}{$\rm{F_{\nu}^{\lambda}}$ \textbf{[$\mu$Jy]}}			&	&	\multicolumn{2}{c}{$\rm{F_{ALMA, rms}}$ \textbf{[$\mu$Jy]}}		\\ \cline{5-6}  \cline{8-9}
		&							&						&		(\%)	 	&	$\rm{F_{12\mu m}=150\,mJy}$	&	$\rm{F_{12\mu m}=10\,Jy}$	& & 1$\sigma$ 1hr							&					1$\sigma$ 10hr 			 \\ \hline
Torus	&	100GHz/3000$\rm{\mu m}$	&	$-$9.33 [$-$10.68,$-$7.93] &	0.02 [0, 10.8]	&	$<$0.1				&	$<$10		& &	  12			&	4			\\
		&	353GHz/850$\rm{\mu m}$	&	$-$6.18 [$-$7.46,$-$4.86]	&	34.6 [0.1, 99.7]	&	$<$80					&	4 [0, 6400]		& &	85			&	30	 	\\
		&	666GHz/450$\rm{\mu m}$	&	$-$4.6 [$-$5.84, $-$3.31]	&	96.2 [5.5, 99.9]	&	2 [0, 2200]				&	100 [0.2, 160000]		& &	260			&	80	 \\ \hline
Jet		&	100GHz/3000$\rm{\mu m}$	&	$-$5.57 [$-$7.01,$-$4.12]	&	99.98 [89.2, 100.0]	&	$<$100				&	20 [0, 8200]		& &		12			&	4 \\
		&	353GHz/850$\rm{\mu m}$	&	$-$5.89 [$-$7.34, $-$4.44] &	65.4 [0.3, 99.9]	&		$<$60				&	7 [0, 4400]		& &		85			&	30 \\
		&	666GHz/450$\rm{\mu m}$	&	$-$6.06 [$-$7.51, $-$4.61] &	3.8 [0, 94.5]	&		$<$40				&	5 [0, 3200]		& &		260			&	80 \\
\hline	\hline
\end{tabular}
\end{center}
\caption{Expected sub-mm versus X-ray flux ratio (Col. 3), percentage (Col. 4), and the flux density [$\mu$Jy] of an hypothetical source with 150mJy (Col. 5) and 10Jy (Col. 6) for the torus and jet contributions. These values are computed at  the three frequencies/wavelengths. The noise level (rms, 1$\sigma$ level [$\mu$Jy]) reached with 1\,hr or 10\,hr of ALMA integration time (IT) is shown in Col. 7  and 8.  }
\label{tab:detecnumbers}
\end{table*}

\subsection{Detectability of the torus using ALMA}
\label{Sec:DetectabilityOfTheTorus}
We discussed in previous sub-sections in which conditions the torus and jet have a maximum or minimum contribution to the sub-mm wavelengths. Here we study (1) the detectability of the torus compared to the jet contribution and (2) the detectability of the torus considering the sensitivity limits of ALMA at the studied frequencies. 

We computed the mean, 10\% and 90\% percentiles of the torus and jet distributions at 100\,GHz, 353\,GHz, and 666\,GHz compared to the X-ray flux ($\rm{F_{\nu}/F_{(2-10keV)}}$). These numbers are reported in Col.\,3 of Table \ref{tab:detecnumbers} (10\% and 90\% recorded within brackets). We then computed the median percentage of contribution for both torus and jet contribution. The minimum (maximum) percentage is computed using the minimum (maximum) ratio for one of the component and the maximum (minimum) ratio for the other. These percentages are reported in Col.\,4 of Table \ref{tab:detecnumbers}. The jet component fully dominates the 100\,GHz/3000\,$\rm{\mu m}$ band. The torus component could dominate over the jet component in the 353\,GHz/\,850$\rm{\mu m}$, although the median contribution is $\rm{\sim}$35\%. As expected, the best chances to detect the torus over the jet contribution are obtained for the 666\,GHz/450\,$\rm{\mu m}$ frequency, with an average $\rm{\sim}$96\% of torus component. However, even using this frequency, the contribution of torus can be as low as $\rm{\sim}$5\%. 

We then computed the expected flux density at sub-mm wavelengths for a typical AGN of $\rm{F_{12\mu m}=150\,mJy}$ and $\rm{F_{12\mu m}=10\,Jy}$ (reported in Cols. 5 and 6 in Table \ref{tab:detecnumbers}). For that purpose we used the X-ray to mid-IR relation to convert mid-IR flux $\rm{F_{12\mu m}}$ to X-ray flux $\rm{F_{(2-10 keV)}}$. We then used the median, minimum and maximum for $\rm{F_{\nu}/F_{(2-10keV)}}$ to estimate the expected flux at the three sub-mm frequencies studied. The ALMA sensitivity limits, or the root mean square (rms), according to the exposure time calculator, considering one hour (on-source) observing time using the most sensitive ALMA configuration and the full bandwidth (BW) receivers ($\sim$ 8 GHz BW) are $\sim$12, $\sim$85, and $\sim$260\,$\mu$Jy at 100\,GHz/3000\,$\rm{\mu m}$, 353\,GHz/850\,$\rm{\mu m}$, and 666\,GHz/450\,$\rm{\mu m}$, respectively. Comparing our results with these limits, further constrains the plausible detectability of the torus. A torus at 100\,GHz/3000\,$\mu$m is not detectable, even with observations of 10 hours long. This integration time would gives an rms of 4$\mu$Jy, but it is still not enough to detect, with a marginal detection limit (5$\sigma$), the flux density of the brightest torus (10 $\mu$Jy at 100\,GHz/3000\,$\rm{\mu m}$). Using the 350\,GHz/850\,$\rm{\mu m}$ band it is possible to have a good detection (a 10\,$\sigma$ detection) for a torus with flux density larger than 850\,$\mu$Jy ($\sim$ 1\,mJy), selecting a sample of bright AGN (with flux density as bright as $\sim$ 10\,Jy). Fainter tori are not detectable even with larger integration time. A 10-hr, integration time observation at this frequency band can reach an rms of 30\,$\mu$Jy. However, this is still not enough to detect a 4\,$\mu$Jy torus or the tori of those AGN with hundred of mJy flux density. Therefore, although a large fraction of torus could be detected at the 353\,GHz/850\,$\rm{\mu m}$ (see col. 4 in Table \ref{tab:detecnumbers}), its median values would never be detected for an AGN with 150\,mJy at 12\,$\rm{\mu m}$ and for an AGN as bright as 10\,Jy. 

A larger flux density is expected for the torus contribution at the 666\,GHz/450\,$\rm{\mu m}$.  At this frequency it is possible to have good detection (a 10 $\sigma$ detection) for tori brighter than $\sim$2.6 mJy, therefore selecting bright tori of bright AGN (with flux density as bright as $\sim$ 10 Jy). Using longer integration time, e.g. 10\,hr, the rms can go down to $\sim$ 80\,$\mu$Jy, giving the chance to detect tori brighter than 800\,$\mu$Jy. However, in this scenario it is still difficult to detect the flux density of the brightest torus for a 150\,mJy AGN (it would be detected at $\sim$ 3$\sigma$ detection limit).  Overall, the result is that, even if on average the torus dominates at this frequency (see col. 4 in Table \ref{tab:detecnumbers}), it will not be detected considering the median of the distribution. 

Knowing the ALMA rms estimate of 1\,hr or 10\,hr integration time, the median value of the radio jet is, in principle, not detected in the sub-mm range. However, bright radio jets for bright AGN (with flux density as bright as $\sim$ 10 Jy) can contaminate the sub-mm continuum flux density at all the three frequency/wavelength bands considered.

Thus, only the brightest, largest, thickest, and with the largest number of clouds tori (see Section \ref{sec:toruscontrib}) will produce enough flux to be detected at either the 353\,GHz/850\,$\rm{\mu m}$ and 666\,GHz/450\,$\rm{\mu m}$ sub-mm frequencies. To give a general estimate of the ALMA detectability of the torus and to summarize the above paragraph, we can say that 1\,mJy torus (characteristics of bright AGN, with flux density at 12 $\mu$m S $\sim$10 Jy) can be detected with 10-$\sigma$ detection limit, observing:  1\,hr at 353\,GHz/850\,$\rm{\mu m}$ and 10\,hr at 666\,GHz/450\,$\rm{\mu m}$. However, the observer has to be aware of the possible contamination (and in some cases with very high percentage) by the radio jet at these sub-mm bands.  In fact, Tab.\ \ref{tab:detecnumbers} shows that the median value of the jet contributor, is not detected at any the three ALMA frequencies and considering 1\,hr or the longer 10\,hr integration time. However, bright radio jet for bright AGN might dominate at these frequencies.

\section{Torus detection with ALMA through observations}\label{sec:data}

\subsection{Cases of study}\label{sec:casesofstudy}

With no attempt to define a complete sample, we select four AGN to confront our predictions on the detectability of the torus at sub-mm wavelengths with actual data. As we showed in section \ref{sec:theory}, the detectability is sensitive to the brightness at X-ray/mid-IR wavelengths, the accretion rate, and the BH mass. It also depends on the torus parameters but we cannot know these parameters ahead of the analysis. 

We select targets with available radio data to trace the jet contribution, mid-IR data to trace the torus contribution, and with continuum sub-mm observations. Additionally, we selected our targets by probe different optical classes. These targets are: the Low-Luminosity AGN (LLAGN) NGC~1052, the Type-2 Seyfert NGC~1068, the Type-1.5 Seyfert NGC~3516, and the QSO IZw~1. Our four AGN cover a wide range of X-ray luminosity with more than 2 orders of magnitude, a factor of 10 in BH masses, and a wide range of 12 $\rm{\mu m}$ fluxes from 150\,mJy to 35\,Jy. General information about our selected targets are shown in Table \ref{tab:ObsParam}. 

{\bf NGC~1052} is a nearby ($z=0.005$; we used the average distance independent on redshift reported in NED\footnote{The NASA/IPAC Extragalactic Database (NED) is operated by the Jet Propulsion Laboratory, California Institute of Technology, under contract with the National Aeronautics and Space Administration.}) elliptical galaxy which hosts a LLAGN in its center \citep[with luminosity between 1GHz and 100 GHz: L$\rm{_{1-100GHz} = 4.4 \times 10^{40}}$ erg s$\rm{^{-1}}$ ;][]{Wrobel1984}. It shows a twin-jet system in the east-west direction at radio kpc and pc scales, oriented close to the plane of the sky \citep{Vermeulen2003,Cooper2007} which is contained within the optical galaxy. Its optical spectrum is characterized by strong forbidden lines from low-ionization states. For this reason this source is considered a prototypical LINER \citep[low-ionization nuclear emission line region; ][]{Heckman1980} galaxy. The X-ray images of NGC~1052 show a point-like X-ray source and its X-ray spectrum is extremely flat, most likely explained by the advection dominated accretion flow (ADAF) mechanism \citep{Guainazzi2000a}. To model the AGN X-ray spectrum, large absorbing column densities have been discussed, supporting the idea of a highly dense obscuring torus \citep[e.g., ][]{Risaliti2002}. Evidence of an obscuring torus has also been suggested from VLBI observations in the radio regime; a prominent emission gap has been detected between the twin-jet system \citep[e.g., ][]{Kadler2003}. Although NGC~1052 has been part of large samples to study the torus component in LLAGN \citep{Mason12,Gonzalez-Martin15,Gonzalez-Martin17}, its torus has never been modeled individually.

{\bf NGC~1068} ($z=0.002$; redshift-independence reported in NED) is the prototype Seyfert 2 galaxy, where the central engine is supposed to be blocked by the dusty torus. It has been studied at a large number of wavelengths: e.g., in the radio band (at 5-GHz and 8.4 GHz) using Very Long Baseline Interferometry (VLBI) technique \citep{Muxlow1996,Gallimore2004}; in mm band \citep{Krips2006}; in near- and mid-IR bands \citep{Marco2000,Galliano2002, Jaffe2004,Galliano2005}; and at high energy X-rays \citep{Guainazzi2000New}. The relativistic jet is prominent and it extends for several kpc in both directions. It changes the direction about 0.2$\arcsec$ from the nuclear region. This change is presumed to be the result of an interaction with a molecular cloud \citep[][]{Gallimore2004}. Significant near- and mid-IR emission is associated with the inner radio jet \citep{Marco2000, Jaffe2004, Galliano2005} which is presumed to be the result of shock heating of the dust in the ISM by the passage of the radio jet. \cite{Lopez-Rodriguez18} used SOFIA, infrared and sub-mm observations in order to characterize the emission and distribution of the dust in NGC~1068. They fitted the nuclear SED of NGC~1068 using clumpy and smooth torus models, finding for clumpy torus: an angular width $\rm{{\sigma=43^{+12}_{-15}}^{\circ}}$, a radial thickness $\rm{Y=18^{+1}_{-1}}$, a number of equatorial clouds $\rm{N_{0}=4^{+2}_{-1}}$, a radial distribution $\rm{q=0.08^{+0.19}_{-0.06}}$, an optical depth $\rm{\tau_{V}=70^{+6}_{-14}}$,  a viewing angle $\rm{i={75^{+8}_{-4}}^{\circ}}$,  an inner radius $r_{in} = 0.28^{+0.01}_{-0.01} pc$, and an outer radius $r_{out} = 5.1^{+0.4}_{-0.4} pc$ ; and for the smooth torus: an opening angle $\rm{\theta_{A}=37^{+23}_{-8}}$, a radial thickness $\rm{Y=20^{+4}_{-4}}$, a fixed radial distribution $\rm{q=1}$, an optical depth $\rm{\tau_{V}=250^{+20}_{-10}}$, a viewing angle $\rm{i={79^{+7}_{-10}}^{\circ}}$, an inner radius $r_{in} = 0.41^{+0.05}_{-0.02} pc$, and an outer radius $r_{out} = 8.5^{+7.9}_{-0.7} pc$. Moreover, \cite{Fritz06} fitted the SED of NGC\,1068 using their smooth model and their parameters were: an opening angle $\rm{\theta_{A}=160^{\circ}}$, a radial thickness $\rm{Y=20}$, a radial density distribution $\rm{\beta=-1}$, a altitude density distribution $\rm{\gamma=6}$, an optical depth $\rm{\tau_{V}=8}$ and a viewing angle $\rm{i=70^{\circ}}$.   They obtained a minimum radius $r_{min} = 0.82 pc$, and a maximum radius $r_{max} = 16.4 pc$. Note that with both models, clumpy and smooth, a similar torus size is obtained, given by the $Y$ parameter.

{\bf{NGC\,3516}} is a bright type 1.5 Seyfert galaxy. It has a redshift of z = 0.012 (using redshift-independent measurement reported in NED). Its SMBH is estimated to have a mass of M$_{BH}$=1$\times$10$^{7}$M$_{\odot}$ \citep{Onken03}. NGC\,3516 has been extensively observed and studied at UV and X-rays. This source has strong UV absorption lines, specifically NV, CIV, and SiIV to have been detected with the International Ultraviolet Explorer \citep[IUE, ][]{Ulrich83}. In several works it has been found that these lines vary on timescales as short as weeks \citep{Voit87,Walter90,Kolman93}. \citet{Kraemer02} found optical absorption features associated to eight distinctive kinematic components. This complex outflowing features can also be seen at X-rays  \citep{Netzer02,Turner05,Huerta14}. Esparza-Arredondo et al. (in prep.) studies the mid-IR and X-ray observations of NGC\,3516 to explored the torus parameters. They fitted the SED using the mid-IR \citep[smooth torus model by ][]{Fritz06} and X-ray \citep[Borus model, ][]{Balokovic18} models simultaneously and found that the torus parameters for this sources are: viewing angle $\rm{i = 10^{\circ}}$, torus angular width $\rm{\sigma_{torus}}$ = 60$\rm{^{\circ}}$, dust density radial profile $\rm{\gamma}$=5.9, $\rm{\beta= 10^{-2}}$, a radial extent $\rm{Y = 11.9}$, and an optical depth $\rm{\tau_{\nu}=3.5}$. No direct observations of its radio jet have been performed so far. 



{\bf{I\,Zw1}} is on of the prototypical Narrow Line Seyfert 1 \citep{Osterbrock1985} with narrow permitted lines, weak [O III] emission, strong [Fe II] emission, a steep soft X-ray spectrum, and strong X-ray variability \citep{Sargent1968,Phillips1976,Oke1979,Boller1996,Veron-Cetty2004,Gallo2004}. It is also classified as an infrared-excess Palomar Green QSO and a possible candidate for an ongoing minor merger. \citep{Schmidt1983}. Its redshift is z = 0.059 \citep{Ho2009}, therefore I\,Zw 1 is considered one of the closest QSOs. \cite{Martinez-Paredes17} studied the near- and mid-IR continuum emission of I\,Zw1, fitting the SED to Clumpy models using BayesClumpy \citep{AsensionR09}. The parameters found were: a viewing angle $\rm{i=79 \pm 3^{\circ}}$, a torus angular width $\rm{\sigma_{torus}<17^{\circ}}$, a radial extent $\rm{Y=67^{+15}_{-12}}$, a number of equatorial clouds $\rm{N_{0} < 4}$, a radial distribution $\rm{q= 1.56^{0.04}_{0.03}}$, and an optical depth $\rm{\tau_{\nu} = 54 \pm 4}$. No direct observations of its radio jet have been performed so far.





\begin{table*}[!t]
\def\arraystretch{1.1}
\caption{General information of our selected case of study.}
\begin{center}
\tiny
\begin{tabular}{lccccccccc}
\hline \hline
Obj name & 	Coordinates &  Distance 	& $\rm{\log(M_{BH})}$ 	& $\mathrm{L_{(2-10) keV}}$ & $\rm{F_{12\mu m}}$ & Class. & \multicolumn{3}{c}{Observations} \\ \cline{8-10}
   		& [J2000]	&(Mpc) 	 		&  ($\rm{M_{\odot}}$) 	& (erg/s)  					& (mJy) 			&   				&  \multicolumn{1}{c}{mid-IR}& cm &sub-mm  \\
(1) & (2) & (3) & (4) & (5) & (6) & (7)  & (8)  & (9) & (10) \\
\hline
NGC~1052	& 02:41:04.798 \ --08:15:20.75 	&  20.6 &  $\rm{8.1\pm0.3^{(1)}}$  & $\mathrm{4.60 \times 10^{41(a)}}$  & 150 & LINER   & IRS(LR)/\emph{Spitzer}  & JVLA & ALMA \\
			&				&		&			&		&		&		&		&	Culgoora-3-3	& \\
			&				&		&			&		&		&		&		&	11-m NRAO	& \\
NGC~1068	& 02:42:40.711 \ --00:00:47.81	& 10.6  & $\rm{6.9\pm0.1^{(2)}}$ & $\mathrm{1.0 \times 10^{41(b)}}$ & 17$\rm{\times 10^{3}}$ &  Seyfert 2  & T-ReCS/Gemini  & JVLA& ALMA \\
NGC~3516	& 11:06:47.490 \ +72:34:06.88	& 51.5   & $\rm{7.0\pm0.3^{(3)}}$  & $\mathrm{2.51 \times 10^{43(c)}}$  & 300 &  Seyfert 1.5 & IRS(HR)/\emph{Spitzer}  & JVLA& CARMA\\
 I Zw 1		& 00:53:34.940 \ +12:41:36.20	& 240.7  & $\rm{7.13^{(4)}}$ & $\mathrm{7.1 \times 10^{43(d)}}$  & 430 &  QSO & IRS(LR)/\emph{Spitzer}  & JVLA& ALMA \\
			&				&		&		&		&		&		&		&		& JCMT \\
\hline\hline
\end{tabular}
\end{center} 
\label{tab:ObsParam}
\tablecomments{Observational details of our selected objects. Col.\,1: object name,  Col.\,2: coordinates in J2000, Col.3: distance measured in Mpc, Col.\,4: SMBH mass in logarithmic scale, Col.\,5:  $\mathrm{2-10 \ keV}$ X-ray luminosity measured in $\mathrm{erg \ s^{-1}}$, Col.\,6: observed 12$\rm{\mu m}$ flux, and Col.\,7: the AGN classification. References for BH masses: (1) BH mass determined using the correlation between stellar velocity dispersion (from HyperLeda) and BH mass (1) \citet{Lorena-2014}; (2) \citet{Lodato03}; (3) \citet{Onken03} and (4) \citet{Crummy06}. References for X-ray luminosities: (a) \citet{Brenneman09}; (b) \citet{Capi06}; (c) \citet{Edelson99} and (d) \citet{Zhou_Zhang10}. Note that no error was reported by \citet{Crummy06} for the BH mass of IZw1.}
\end{table*}

\subsection{The data}

\subsubsection{Mid-IR data}

The sources NGC~1052, NGC~3516, and IZw1 are point-like sources at
several wavelengths \citep[e.g][]{Schmidt83, Burtscher13, Capetti05},
and their mid-IR emission is mostly dominated by the AGN itself
\citep[Esparza-Arredondo et
al. in prep.,][]{Gonzalez-Martin17, Martinez-Paredes17}. For this reason we used the high and low-resolution
\emph{Spitzer}/IRS spectral data from the
CASSIS\footnote{http://http://cassis.sirtf.com/atlas/ } catalog
\citep[the Cornell AtlaS of \emph{Spitzer}/IRS Sources,][]
{Lebouteiller11}. The resolution of \emph{Spitzer}/IRS is R $\rm{\sim}$
60-130. CASSIS provides flux-calibrated nuclear spectra associated with
each observation. The source NGC\,1068, is characterized by well-known mid-IR extended emission that dominates the Spitzer spectrum \citep[see ][]{Lopez-Rodriguez18}; therefore high angular resolution
data are mandatory to decontaminate the mid-IR emission from the
contribution of the host galaxy. For this reason, we used the N and Q
bands data from the Thermal-Region Camera Spectrograph \citep[T-ReCS, Telesco
et al. 1998,][]{ DeBuizer05} located in the 8.1 m Gemini-South
Telescope. These data better isolate ($\rm{\sim}$0.3-0.4 $\arcsec$ PSF)
the torus component from other dust contributors. T-ReCS data were
processed using the pipeline RedCan \citep{Gonzalez-Martin13} and have
also been previously used by \citet{Ramos-Almeida14}.

\subsubsection{Radio cm and sub-mm data}

For the purpose of our project, we collected radio continuum data observed with the Karl. G. Jansky Very Large Array (JVLA) of the National Radioastronomy Observatory (NRAO)\footnote{The NRAO is a facility of the National Science Foundation operated under cooperative agreement by Associated Universities, Inc.} in its A-configuration to be able to isolate the radio emission of the the central engine as much as possible. We obtained the A-configuration radio data from the JVLA data archive or literature. For the sources I\,Zw 1 and NGC\, 3516, in order to obtain a better radio SED fit, we also added data from literature obtained at different JVLA configuration. These data points continue to well represent the central engine flux density. As well for the source NGC\,1052, two data points obtained with radio instruments different from the JVLA, have been used to better constrain the radio SED. Information about the JVLA project ID, the JVLA A-configuration clean beam, its position angle (PA) and literature references, are reported in Table \ref{radioinfo} in Appendix \ref{RadioInfoAppendix}.

In the sub-mm window, we collected data from the ALMA archive. When ALMA data were not available, sub-mm flux densities obtained using other sub-mm instruments and already published, have been used. In particular, we used the Combined Array for Research in Millimeter-wave Astronomy (CARMA) data published by \cite{Behar2018} for the source NGC\,3516 and the James Clerk Maxwell Telescope (JCMT)  data published by \cite{Hughes1993} for the source I\,Zw1. Table \ref{sub-mminfo}, in Appendix \ref{RadioInfoAppendix}, summarizes the sub-mm information at our disposal. 


JVLA archival data have been calibrated using the data-reduction package CASA (Common Astronomy Software Applications \footnote{https://science.nrao.edu/facilities/vla/data-processing}; version 5.1.2). Standard procedure for flux density, using the standard calibrators 3C\,286 and 3C\,48, and phase calibration have been applied. Moreover, for the more recent wide band projects (IDs: 16B-289 and 16B-343) also channel flagging procedures and standard bandpass and delay calibration have been performed. 
From the calibrated data, Stokes I images were made for all the targets at each band running the CASA \textit{clean} task. In order to isolate as much as possible the central radio component, we performed the cleaning using \textit{robust 0} or \textit{uniform} weighting. Because the majority of the data sets come from the old JVLA correlator (consisting on 2 intermediate frequencies, IFs, of 50 MHz bandwidth each), the \textit{nterm} parameter has been set equal to 1 (i.e.,\ the change in the spectral index within that small bandwidth is neglected). On the cleaned maps, a Gaussian fit has been performed on the point like central radio component. Because of the complex morphology of the source NGC\,1068, an additional process have been perform to the radio images at X and Ku bands (8.4 and 15.0 GHz). We convolved the uniform weighted maps to the L band resolution (1.4 GHz with $\sim$ 1$\arcsec$ angular resolution) in order to perform a proper Gaussian fit to the elongated structure resulted from the cleaning. All the compact radio components that are present in the central position of this AGN are well resolved only when reaching sub-arcsec angular resolution (therefore at Q band, 43 GHz, for our case). In order to match spatial resolutions during the fitting procedure, we sum the flux densities of the four 43 GHz radio components (NE, C, S1 and S2) that are clearly resolved and studied by \cite{Gallimore1996} and \citep{Cotton2008}.

ALMA continuum flux densities have been collected analyzing the products continuum Stokes I maps available for each projects at the ALMA archive web page \footnote{https://almascience.eso.org/alma-data/archive}. On the continuum Stoke I maps, a Gaussian fit on the central sub-mm component has been performed to obtain the sub-mm flux density value. The flux densities for each targets at each frequency are listed in Table \ref{FluxDensityAGN} in Appendix \ref{RadioInfoAppendix}.


	\subsection{Fitting procedure and results}
  
\begin{table}
\def\arraystretch{1.1}
\caption{Fitting results of the mid-IR spectra to the smooth torus model described by \cite{Fritz06}.}
\begin{center}
\scriptsize 
\begin{tabular}{ccccc}
\hline\hline 
 Par. 		& \multicolumn{4}{c}{Object}  \\
 					&  NGC\,1052 					& NGC\,1068 					& NGC\,3516 				& I\,Zw1 						\\
\hline 
$i$ 				& $ 61_{-2}^{+4} $ 				& $60\pm9 $ 					& $1.9_{-0.7}^{+1.1}$ 		& $<11$ 						\\ 
$\sigma$ 			& $> 20.0$ 						& $20\pm 8$ 					& $58.5_{-0.4}^{+0.3}$ 	& $>58$ 						\\
$\gamma$  		& 0.14* 							& 0.01* 							& $ < 6.0$ 					& $3.2\pm0.4$				\\
			  		& 	 								& 3.91 							& 		 						&								\\
$\beta$ 			& $ -0.03_{-0.07}^{+0.02} $ 	& -0.99  							& $<0.01$ 					& $-0.95\pm0.04$ 			\\ 
$Y$ 				& $ 10.7\pm0.3 $ 				& $142\pm19 $ 					& $11.5 \pm 0.1$ & $19^{-1}_{+2}$ 			\\
$\tau$  			& $ 4.0_{-0.4}^{+0.5} $ 		& $1.81\pm0.2 $  	& $3.81_{-0.06}^{+0.07}$ 	& $1.37^{-0.12}_{+0.76}$ 	\\
		  			& 						 			& $0.29\pm0.1 $ 		&  						 		& 								\\
$R_{out}$			&  $\sim$0.4 & $\sim$2.3  & $\sim$2.9  & $\sim$8.1 \\
$\chi^2/DOF$ 	& 1217.72/993 					& 70.98/262 						& 9575.07/1471 				& 	 721.05/992				\\
 \hline\hline
 \label{tab:MIRparam}
\end{tabular}
\end{center}
\begin{footnotesize}
 NOTE-- $i$ viewing angle toward the torus; $\sigma$ angular width of the torus; $\beta$ inversely related to the amount of absorption; $\gamma$ exponent of the logarithmic azimuthal density distribution; $Y$ ratio between the outer and the inner radius of the torus; and $\tau$ silicate dust contribution. \textbf{The asterisks represent frozen parameters, which could not be constrained.} First and second rows in NGC1068 shows the two values reported for $\gamma$ and $\tau$ parameters for the N- and Q-bands, respectively (see text).
\end{footnotesize}
\end{table}

 \begin{figure*}[!ht]
\begin{center}
\includegraphics[width=1.0\columnwidth]{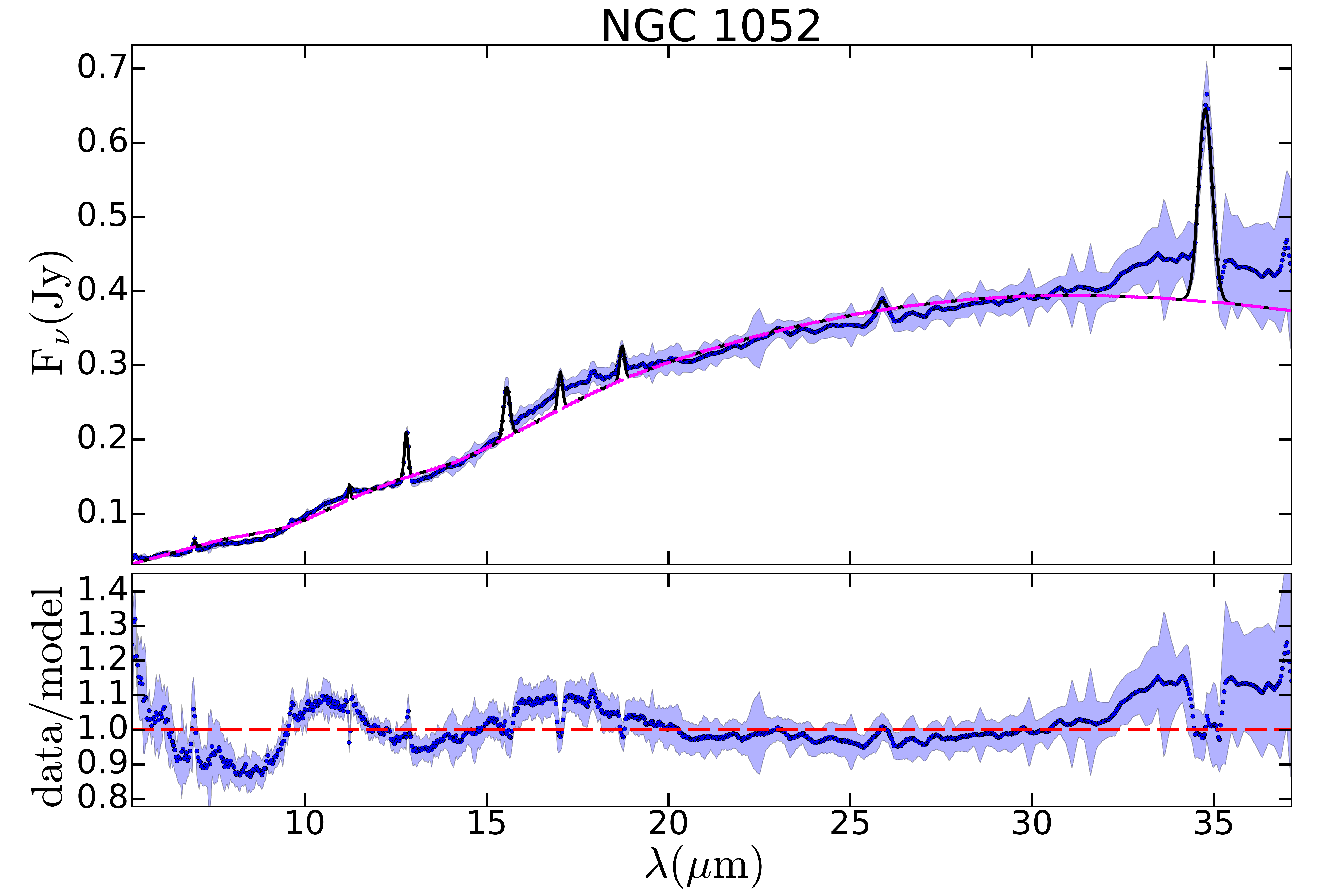}
\includegraphics[width=1.0\columnwidth]{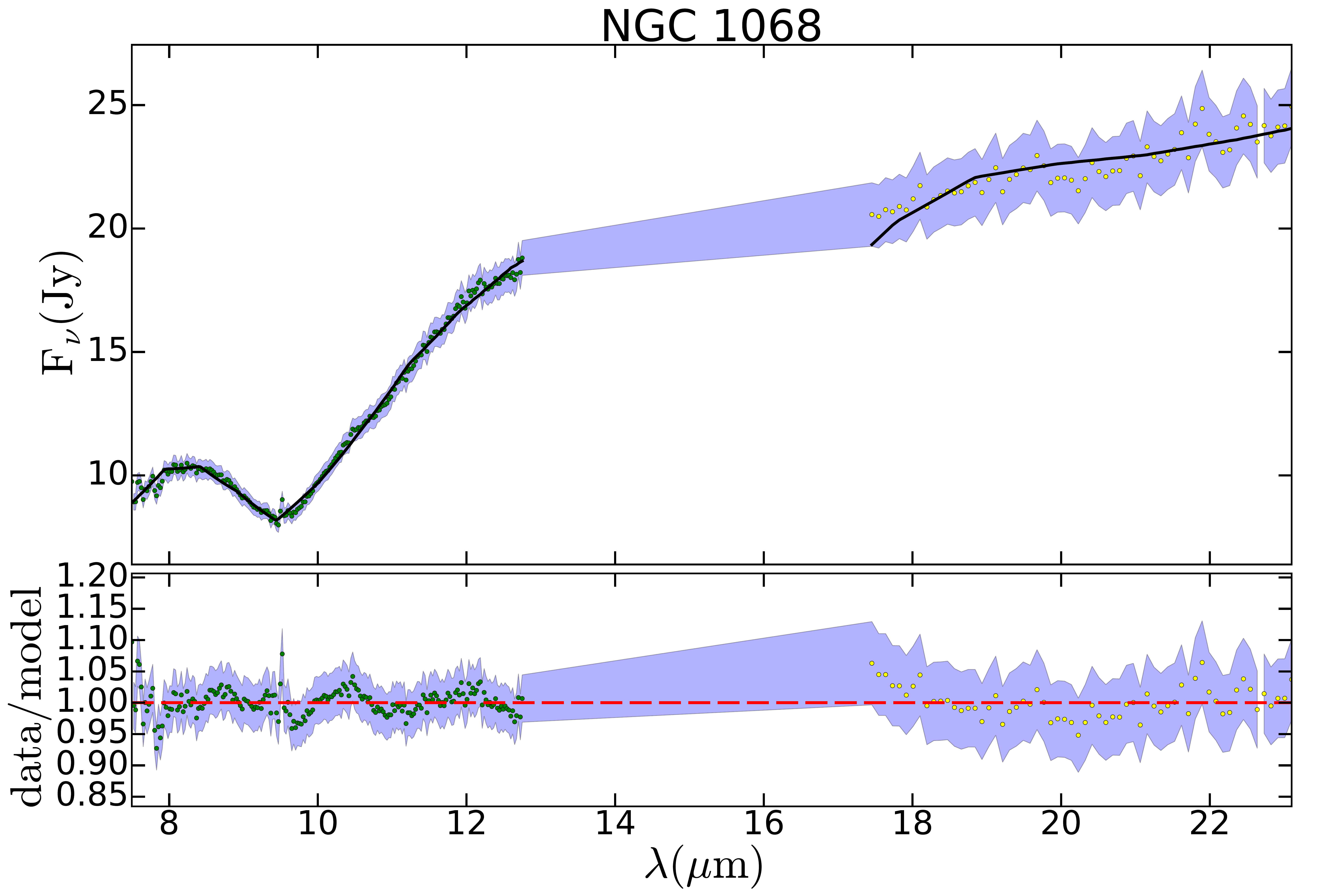}
\includegraphics[width=1.0\columnwidth]{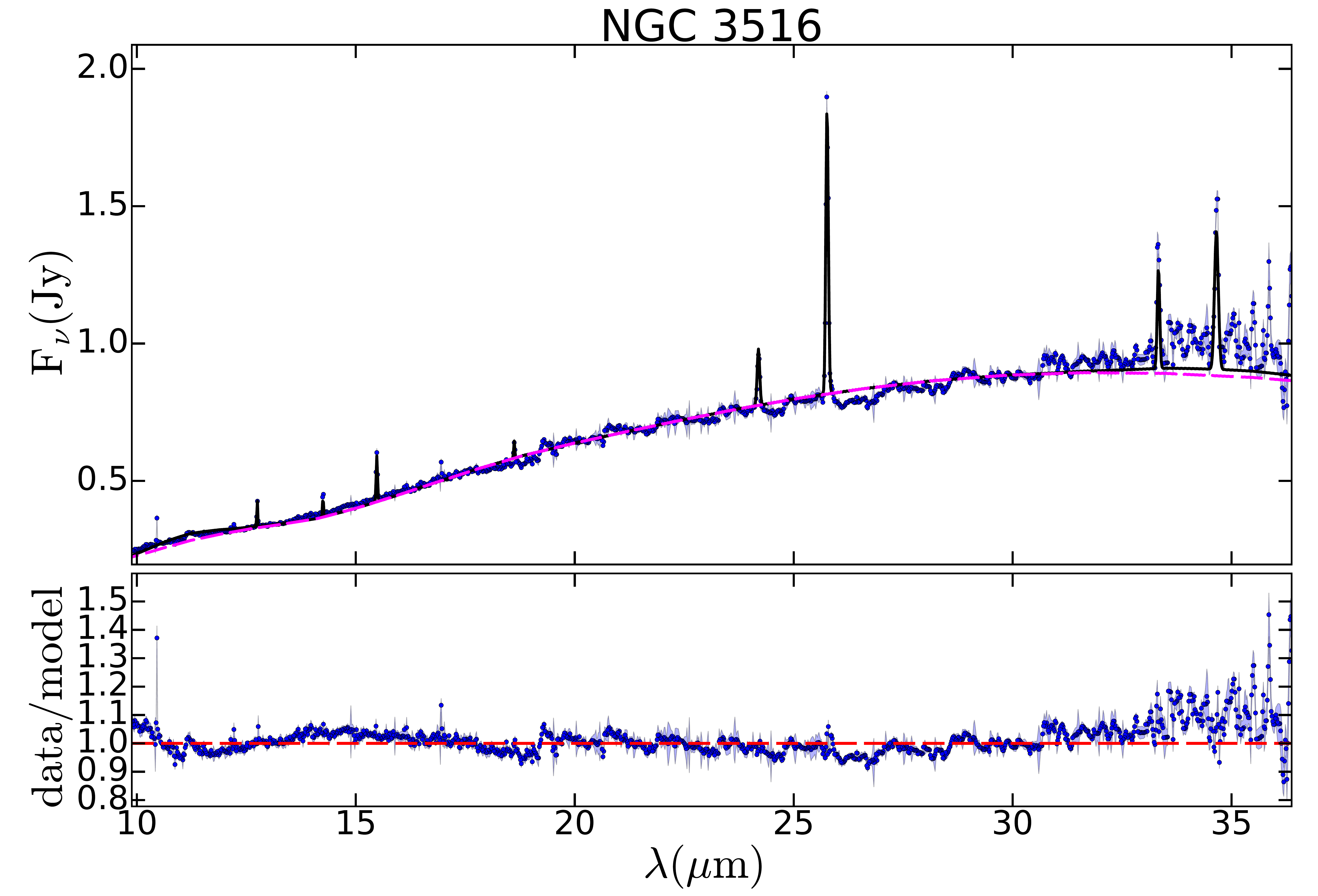}
\includegraphics[width=1.0\columnwidth]{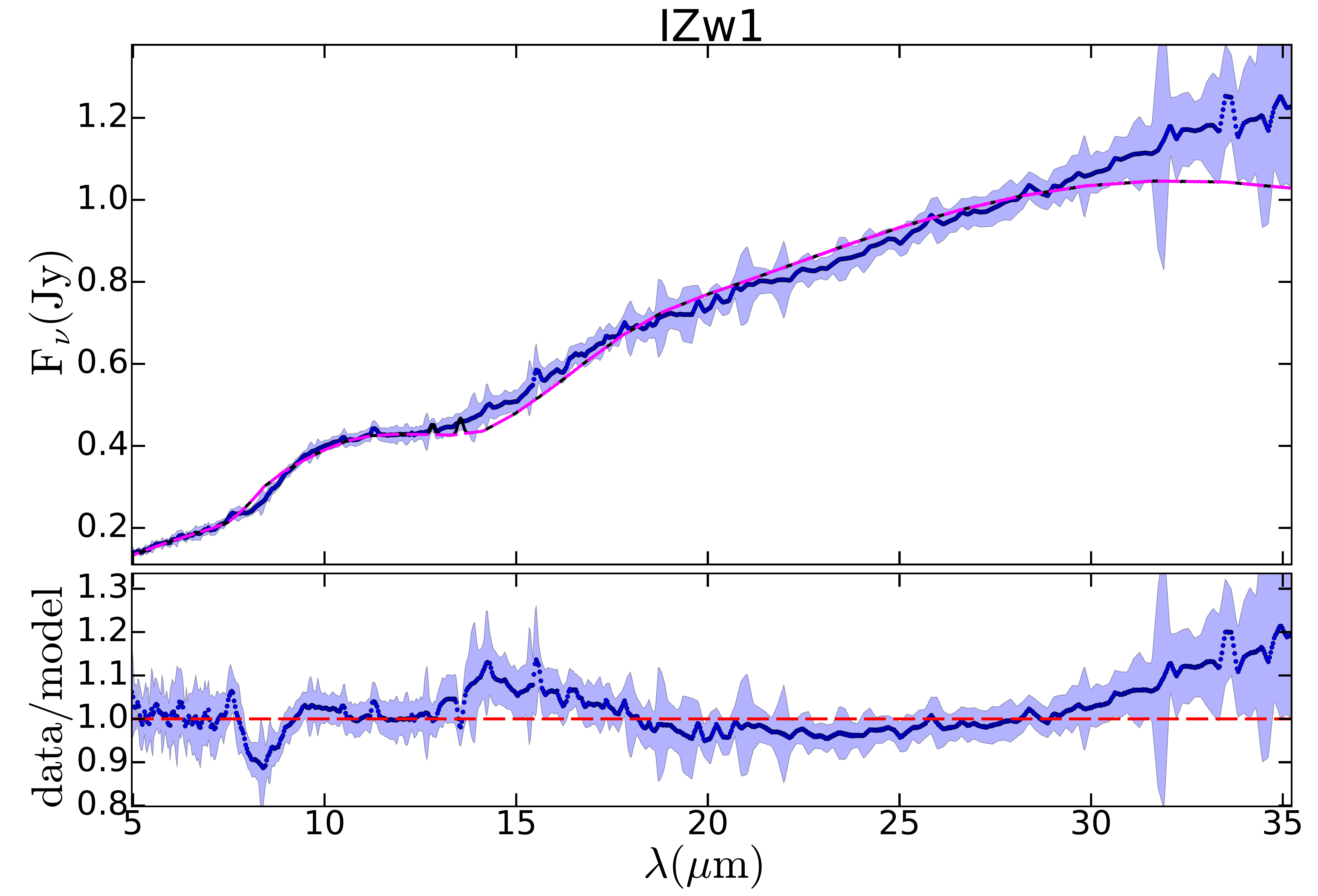}            
\caption{Mid-IR spectral fit to the smooth dusty model described by \citet{Fritz06} for NGC\,1052 (top-left), NGC\,1068 (top-right), NGC\,3516 (bottom-left) and IZw\,1 (bottom-right). For each object we show the best fit to the data in the top panel and the ratio between model and data in the bottom panel. The blue shaded area show the error on the measurement. The red long-dashed line shows the dusty model best-fit and the black solid line show the total fit, i.e. toroidal model $+$ emission lines (see text). } 
\label{fig:MIR_fits}
\end{center}
\end{figure*}     

\subsubsection{Mid-infrared spectral fit to dusty models}\label{sec:mirfit}

Our purpose is to extrapolate the mid-IR spectral fit to sub-mm
wavelengths to establish the detectability of the torus component at
those wavelengths. For that purpose we fit the mid-IR data to dusty
models. Note that we are not interested in well sample the near- to
mid-IR SEDs therefore, we will not focus on the finding of a very detailed derivation of the model
parameters. Instead, we will test our theoretical predictions from model
extrapolations.

The dusty models can be grouped into four classes: smooth \citep{Fritz06,Feltre12}, clumpy \citep{Nenkova08A,Nenkova08B,Hoenig10A,Hoenig10B}, smooth + clumpy \citep{Stalevski12}, and windy \citep{Hoenig17,Siebenmorgen15}. Among them the most extensively used to fit data at mid-IR wavelengths is the clumpy model by \citet{Nenkova08B} due to their large number of SEDs and probed ability to explain the spectra of low luminous \citep{Gonzalez-Martin17}, intermediate luminous \citep{Ramos-Almeida09, Alonso-Herrero11}, and high luminous \citep{Nikutta09, Martinez-Paredes17} AGN. However, smooth models have been proposed to explain some peculiar anomalies of the spectra of high luminous AGN \citep{Martinez-Paredes17}. Furthermore, windy models are based on a more realistic configuration of dust according to dynamical arguments \citep[][and references therein]{Elitzur06,Elitzur09}. We have tested in this work all the models that have public access to a SED library: 1. Smooth model by \citet{Fritz06}; 2. Clumpy model by \citet{Nenkova08B}, 3. Clumpy model by \citet{Hoenig10B}, 4. Windy model by \citet{Siebenmorgen15}, and 5. Windy model by \citet{Hoenig17}. We are currently working on a paper that fully examines the similarities and differences between these five models (Gonz\'alez-Mart\'in et al. 2019A in prep.) and which model better describe current mid-IR spectra of AGN (Gonz\'alez-Mart\'in et al. 2019B in prep.).

We converted these SED libraries to multi-parametric models within the spectral fitting tool {\sc xspec}. {\sc xspec} is a command-driven, interactive, spectral-fitting program within the HEASOFT\footnote{https://heasarc.gsfc.nasa.gov} software. {\sc xspec} provides a wide range of tools to perform spectral fitting to data, being able to process in parallel in order to speed them up. To use these capabilities we need to convert our SEDs to {\sc xspec} format in order to upload our models within {\sc xspec} as additive tables. The basic concept of a table model is that the file contains an N-dimensional grid of model spectra with each point on the grid having been calculated for particular values of the N parameters in the model. {\sc xspec} is able to interpolate on the grid. We have created an additive table for each torus model. For further details on the conversion into {\sc xspec} format please refer to Esparza-Arredondo et al. (in prep.).\\
We also converted the mid-IR \emph{Spitzer} spectrum into {\sc xspec} format using {\sc flx2xsp} task within HEASOFT. This tool reads a text file containing one or more spectra and errors and writes out a standard {\sc xspec} pulse height amplitude (PHA\footnote{Engineering unit describing the integrated charge per pixel from an event recorded in a detector.}) and response files. This allows us to use all the observed data points within {\sc xspec} to fit to the data.

We fitted the mid-IR data for the four sources to the five models described above. In the cases of NGC\,3516 and NGC\,1052 we included some emission lines reported in previous works. We used the \emph{addline} tool available in {\sc xspec} to identify and model each line with Gaussian profiles. All the models provide a reasonable fit to the data. However, the smooth dusty model described by \citet{Fritz06} is the only one that does not need an additional component to provide a good fit to the data. For the others, a stellar component\footnote{The stellar library provided by \citet{Bruzual03} is used to model the stellar component. We have included this SED library as additive tables within {\sc xspec}, similarly to what we did with the torus models.} is needed to account for the short wavelengths (below 10$\rm{\mu m}$). Thus, we used the smooth dusty model reported by \citet{Fritz06} hereinafter since, providing an equally good fit \textbf{($\rm{\Delta \chi^2 / dof < 0.1}$)}, is the simplest baseline model that satisfactory describes the data. The main difference between the resulting fit occurs at short wavelengths, so long wavelengths are quite insensitive to the model used. Therefore we do not expect large discrepancies on the expected contribution of dust at sub-mm wavelengths due to the selection of models. 
\\
In the case of NGC\,1068 we consider initial inclination angles valid for type-2 Seyfert and toroidal sizes according to those found in the literature (see \cite{Garcia-Burillo16} and \cite{Lopez-Rodriguez18}) using mid-IR high angular resolution data. However, we were not able to fit both N- and Q-bands of NGC\,1068 with the same values for the parameters of the torus model. We try different fits; first we link the parameters in both bands, without getting a good fit. We also tried adding the \emph{constant} multiplicative model available in {\sc xspec} to mimic a possible different resolution (or perhaps flux slit losses) between the bands but they failed to reproduce both bands at the same time. Essentially, the shapes of the N- and Q-band spectra cannot be fitted with the same SED. Then we unlink some parameters individually between both bands. We tested several combinations of unlinked parameters (e.g. $Y$ and $\tau$). The best result is obtained unlinking $\gamma$, $\beta$ and $\tau$ parameters (i.e. the coefficients within the density function and the optical depth). Note that although we use the model by \citet{Fritz06} for our fit, we obtain different values of the parameters, since they add to their fit an IR template of a starburst galaxy to reproduce the colder component of the emission of dust. NGC\,1068 has a well reported complex dust distribution \citep[][and references therein]{Lopez-Rodriguez18}. Indeed, the \emph{Spitzer} spectrum shows a continuum flux more than 10 times brighter than the ground-based T-ReCS spectra, indicating external dust contributors at radii larger than a few hundred of parsecs. We interpret that this behavior of the N- and Q-bands reflects a complex structure of the dust within the inner $\rm{\sim}$100 pc that cannot be easily reproduced with a simple dusty model. 

Figure \ref{fig:MIR_fits} shows the best-fit for the targets using the smooth dusty model by \cite{Fritz06}. The resulting parameters of the mid-IR fitting are reported in Table \ref{tab:MIRparam} for completeness purposes and the errors were calculated using the $\mathrm{error}$ tool\footnote{This tool determines the error ranges by sorting the Monte Carlo chain values and taking a central percentage of the values corresponding to the confidence level as indicated by delta fit statistic.} in {\sc xspec} with a 99\% confidence. It is out of the scope of this work to properly characterize the torus properties of our sample. For that purpose, near-IR \citep{Ramos-Almeida14} and/or X-ray (Esparza-Arredondo et al. (prep)) data are needed. 
From our fitting we find that the viewing angle for the type-1 AGN (i.e. IZw1 and NGC\,3516) is consistent with a direct view of the central engine while that of the type-2 AGN (i.e. NGC\,1068) is consistent with the interception of the dusty torus in the line of sight, as previously found using well sampled near to mid-IR unresolved SEDs \citep[e.g.,  Esparza-Arredondo et al. (prep), ][]{Lopez-Rodriguez18,Gonzalez-Martin17,Martinez-Paredes17.} 






\begin{table}
\caption{Best Fit parameters for the radio spectrum fitting}
\label{BestFparams}

\begin{center}
\scriptsize 
\centering
\label{tab:Radiofitting}
\begin{tabular}{llcccc} 
\hline \hline
  Source     & Model & $\alpha_{thin}$ & $\nu_0$  & S$_{\nu_{0}}$ & $\tiny{\chi^2 / \textit{dof}}$ \\
  name    &  &  &  [GHz] &  [mJy]&  \\
\hline
NGC\,1052	&SSA2 	& $-$0.7					&	$\rm{0.06\pm0.01}$ 	&	$\rm{3760\pm430}$		& 515 	\\
									& 			& $-$0.7 					&   $\rm{10.70\pm0.02}$	& 	$\rm{2612\pm14}$		&			\\
NGC\,1068	& PL		& $\rm{-1.07\pm0.01}$	&	$\dots$					&	 		. . .					&  33		\\
NGC\,3516	& SSA		& $\rm{-0.97\pm0.07}$	&	$\rm{1.8\pm0.2}$ 		&	$\rm{6\pm1}$ 			&  0.7	\\
I\,Zw1 			& PL		& $\rm{-0.65\pm0.04}$	&	$\dots$ 					&			. . .					&  0.1	\\
\hline\hline
\end{tabular}
\end{center}
\begin{footnotesize}
NOTE: Col. (1), source name; Col. (2), best fit where PL is a simple power-law fit (fitting the synchrotron optically thin part of the spectrum only), SSA is a single synchrotron self-absorption component, and SSA2 is the combination of two synchrotron self-absorption components; Col. (3), spectral index value in the optically thin part of the spectrum; Col. (4), frequency value where the synchrotron emission changes from optically thick to optically thin and its relative flux density value in Col. (5); Col. (6), the statistical value of $\chi^{2}$/\textit{dof}. 
\end{footnotesize}
\end{table}

\begin{figure*}[ht!] 
	\begin{center}
		\includegraphics[width=1.0\columnwidth]{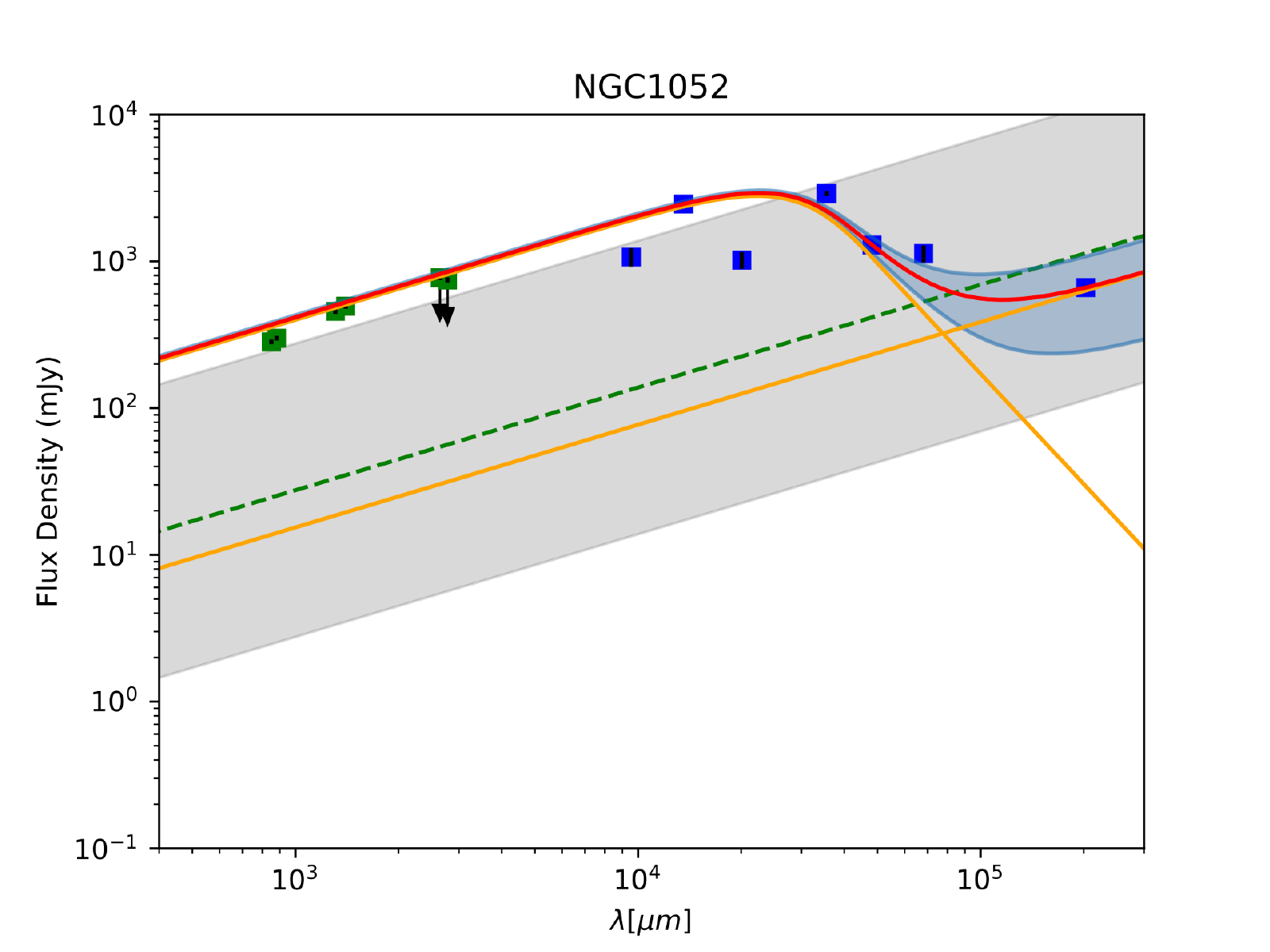}
		\includegraphics[width=1.0\columnwidth]{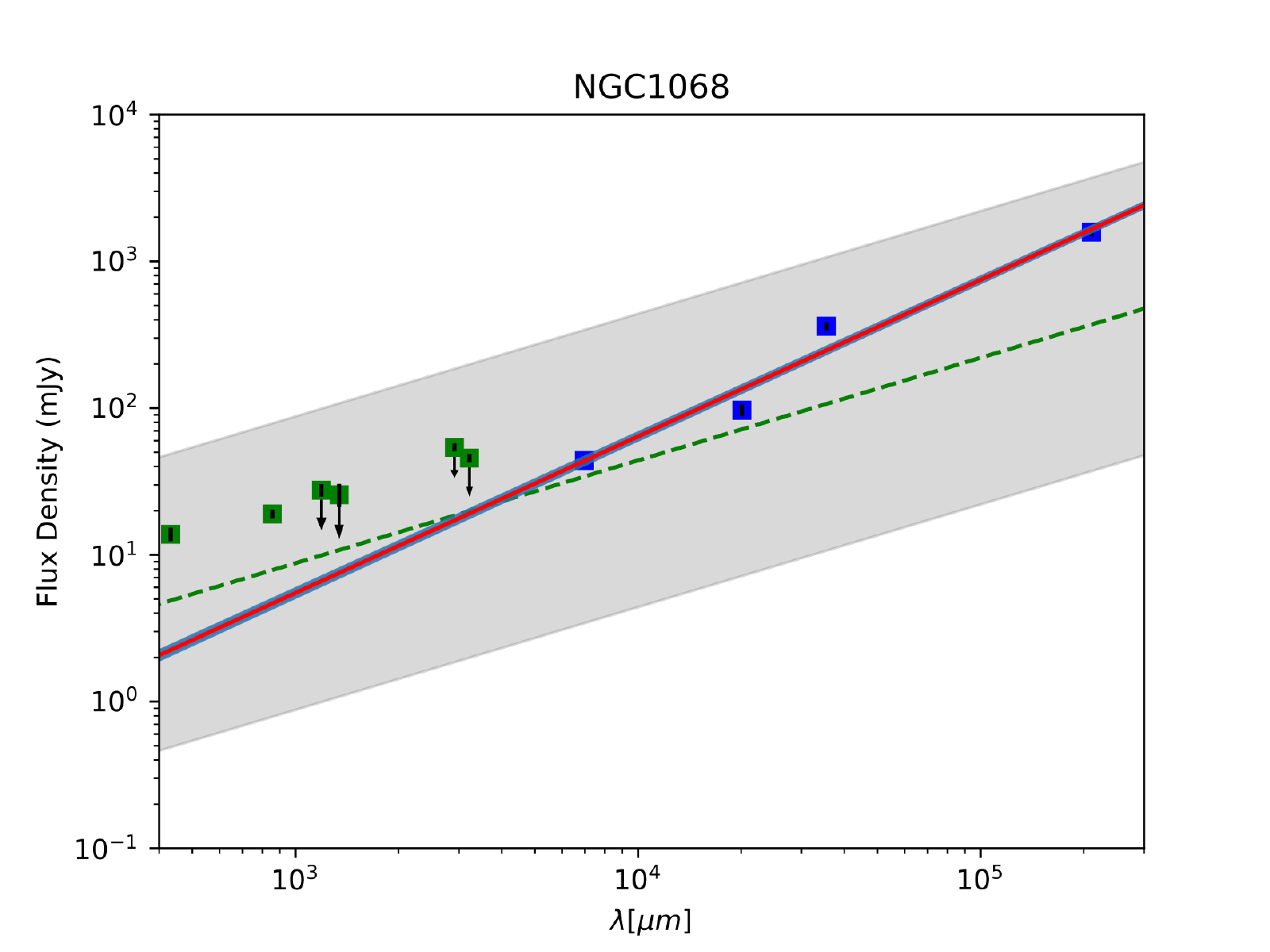}
		\includegraphics[width=1.0\columnwidth]{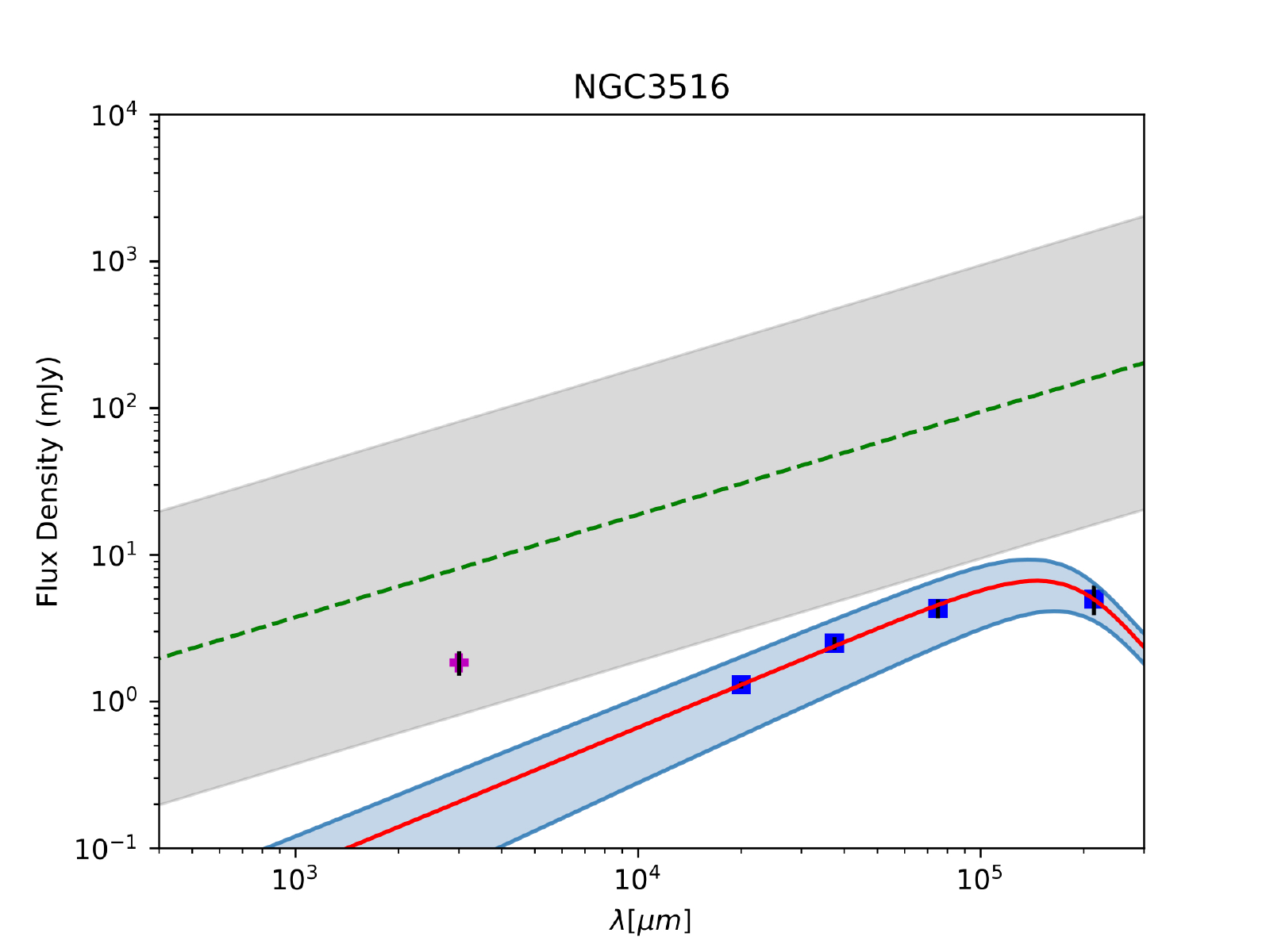}
		\includegraphics[width=1.0\columnwidth]{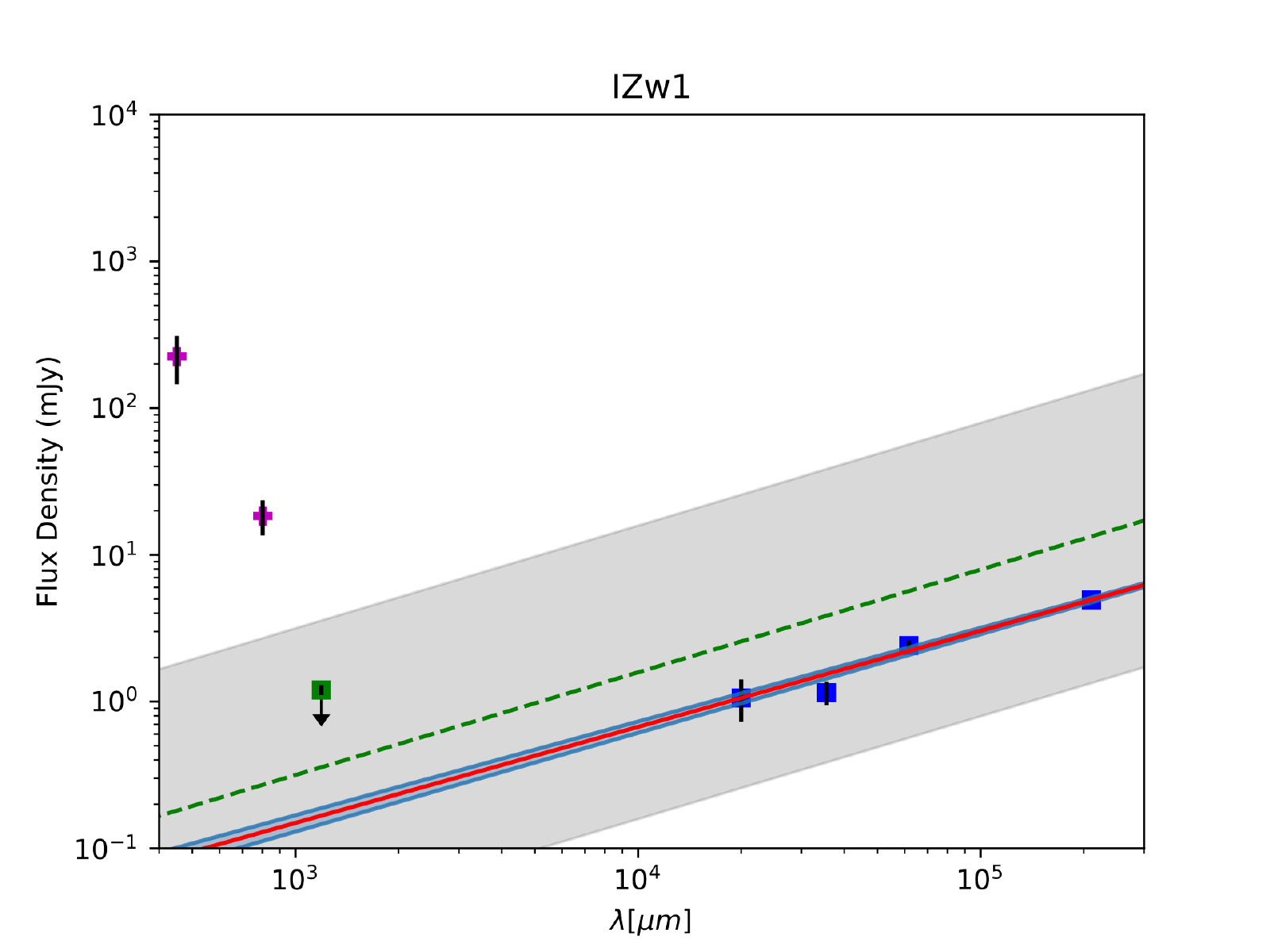}
		\caption{Radio spectrum fitting for the sources NGC\,1052, NGC\,1068, NGC\,3516, and I\,Zw1. Power-law model has been used for NGC\,1068 and I\,Zw1, a single synchrotron component model for NGC\,3516, and two synchrotron components model for NGC\,1052. Blue square points are the cm wavelength data, green square points are sub-mm ALMA data, and magenta crosses are sub-mm data from other instruments. Black arrows mark those data points for which the angular resolution is larger than 0.3$\arcsec$, where extended contributors might be present. The orange straight lines, where present, represent the single synchrotron components while the red straight line and the blue transparent band are the total fit of the radio data (i.e.,\ considering the JVLA/cm data only) together with its error. The fundamental plane prediction together with its uncertainty, of an order of magnitude, is also shown in the plots as the dashed green line and a gray band, respectively. }
		\label{fig:Radiofitting}
	\end{center}
\end{figure*}    

\subsubsection{Radio spectral fit to the synchrotron emission}\label{sec:radiofit}
The total intensity radio spectra of the targets have been fitted using a power law ($\rm{S^{PL}_{\nu}}$), representing the optically thin part of a radio synchrotron spectrum, and a single or a composition of two synchrotron emitting components ($\rm{S^{SSA}_{\nu}}$ and $\rm{^{SSA2}_{\nu}}$), considering homogeneous, self-absorbed sources. Each synchrotron emitting component is modeled with power-law electron energy distributions with spectral index in the optically thick and thin parts of the spectrum $\rm{\alpha_{thick}}$=2.5 and $\rm{\alpha_{thin}}$, respectively. The equations used for the radio spectra fitting are the following:
\begin{itemize}
\item a power law (optically thin synchrotron spectrum) with a slope $\rm{\alpha_{thin}}$ (left to vary):
\begin{equation}
S^{PL}_\nu \propto \nu^{\alpha_{thin}}
;\end{equation}
\item a single synchrotron self-absorption component ($\rm{S^{SSA}_{\nu}}$):
\begin{equation}
S^{ssa}_{\nu} \propto \nu^{2.5}   \left( 1 - \exp\left(- \left(\frac{\nu}{\nu_{0}}\right)^{\alpha_{thin}-2.5}\right)\right)
,\end{equation}
\noindent where $\rm{\nu_0}$ is the frequency where the emission changes from optically thick, with a spectral index of 2.5, to optically thin with a spectral index $\rm{\alpha_{thin}}$ (left to vary);
\item a combination of two synchrotron components ($\rm{S^{SSA2}_{\nu}}$) with fixed $\rm{\alpha_{thin}= -0.7}$ in order to reduce the number of free parameters.
\end{itemize}
\noindent To fit the above functions to the data we used the python task $\textit{curvefit}$ on the cm wavelength data (i.e.,\ JLVA data and cm wavelength from literature) only. For this radio fitting procedure we considered the errors of the data points as absolute. This allows to provide the variance of the estimated fit parameters as the square root of the diagonal of the covariance matrix. For each source, we selected the best model according to the lowest statistical reduced chi-squared test ($\rm{\chi^{2}}$/\textit{dof}) value. 

The resulting parameters of the radio fitting are reported in Tab. \,\ref{tab:Radiofitting}. Fig.\,\ref{fig:Radiofitting} shows the results of the radio fitting of the targets using a single power law for the sources I\,Zw1 and NGC\,1068, one synchrotron component for the source NGC\,3516 and two synchrotron components for the source NGC\,1052. Sub-mm data points (ALMA and CARMA or JCMT) are also shown in the plots (as green squares and magenta crosses respectively). We marked with a black arrow those ALMA data for which the angular resolution is larger that 0.3$\arcsec$ (which correspond to a spatial resolution of $\sim$340, $\sim$140, $\sim$800 pc for NGC\,1052, NGC\,1068 and NGC\,3516 respectively and  $\sim$4\,kpc for I\,Zw 1). These data, together with sub-mm data obtained from other sub-mm telescopes (i.e., magenta cross) could overestimate the flux density at that frequency due to the large beam size. We will discuss this issue in section \ref{sec:discussion}. The radio spectra fitting works very well for the sources I\,Zw1 and NGC\,3516, while the fit shows large $\rm{\chi^2}$/\textit{dof} values for the sources NGC\,1068 and NGC\,1052. The fit cannot represent the JVLA high frequency data point (at 43.4\,GHz) for NGC\,1068 without adding a second synchrotron component in the sub-mm window (see the discussion section). NGC\,1052 shows a complex behavior, that cannot be accounted for with our fit, due to its variability at mid and high radio frequencies (for $\rm{\nu >}$5\,GHz); multi-frequencies instantaneous radio observations could resolve this issue. 

The fundamental plane prediction (green-dashed line and gray band) shown in each plot, gives an idea of the order of magnitude of the flux density of the radio jet. The fundamental plane, considering its uncertainties, follows the trend of the radio emission described by the jet in all cases except for NGC\,3516 where the fundamental plane is an order of magnitude above the radio emission. Moreover, it does not account for the complexity of the radio emission. For example, in the case of NGC\,1052 it is clear that the AGN needs two synchrotron components in order to represent the high radio frequency part of the spectrum (for frequencies larger than $\sim$ 10 GHz or wavelength smaller that $\sim$ 3$\times$10$^4$, see Tab.\,\ref{tab:Radiofitting}); instead, the fundamental plane prediction goes below the high frequency JVLA data. For NGC\,1068 and NGC\,1052, for which we have much more sub-mm values at our disposal, the sub-mm flux densities increase towards larger wavelengths, opposite to what is expected if this emission is associated to dust. Although the possible inaccuracies described above, it is worth to notice that the fundamental plane seems to include, in all the case of study, a non-negligible jet contribution to the sub-mm band.


These fits probe that the sub-mm part of the spectra is still (partially) contaminated by synchrotron emission (perhaps dominating for NGC\,1052 and NGC\,1068). However, a better analysis on this aspect is given in Section \ref{sec:discussion}, where the mid-IR fit and the radio fit are combined together, giving a broad view on the contributors to the SED of these four AGN.

\section{Discussion}\label{sec:discussion}

\begin{figure*}[!t]
\centering
\includegraphics[width=2.\columnwidth]{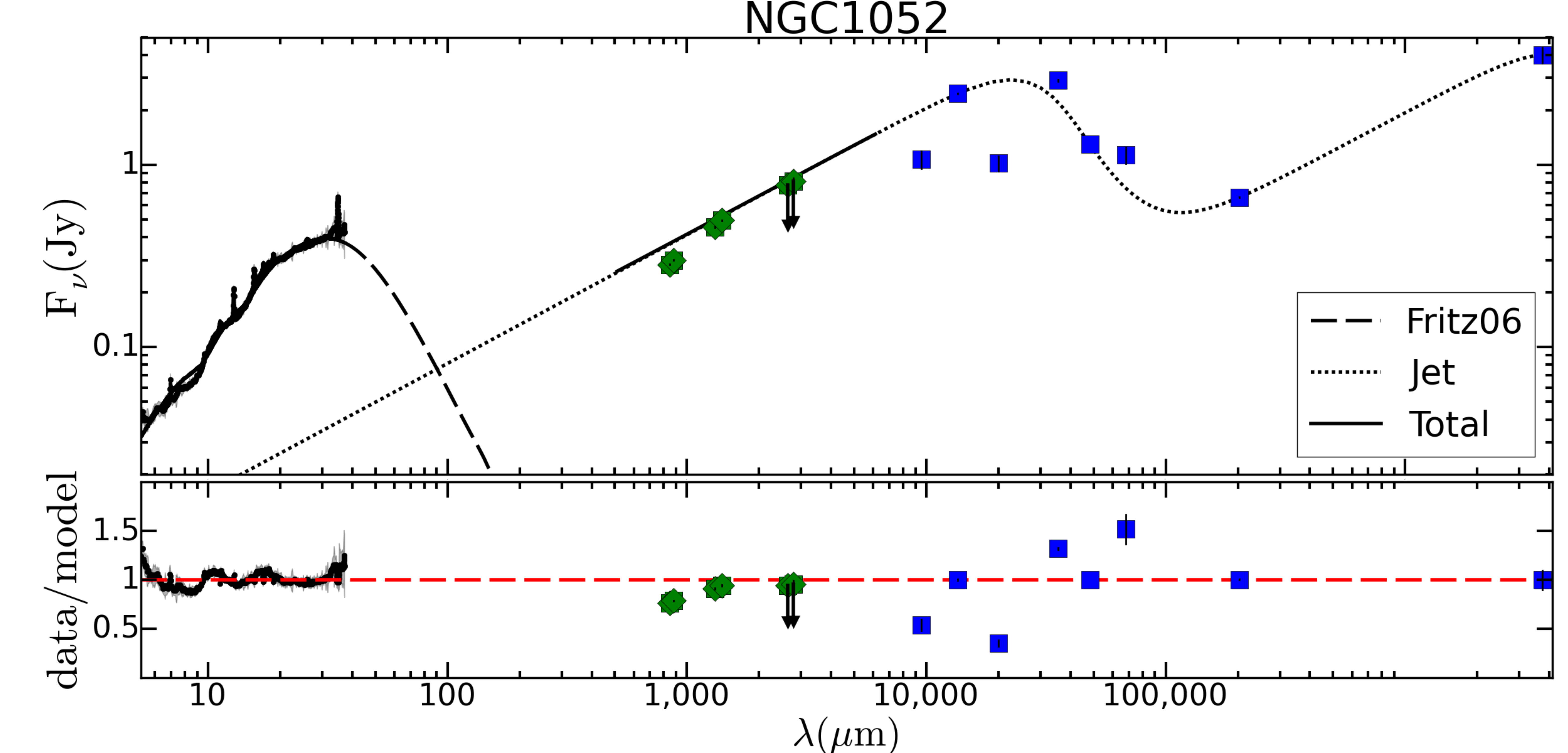}
\caption{SED fit for NGC\,1052 to the mid-IR spectra (black spectrum) with the torus model (long-dashed line) and to the cm wavelengths (blue points) with synchrotron emission (dotted line). Green stars are sub-mm and far-infrared data. Arrows indicate plausible upper limits when the spatial resolution is worse than 0.3$\arcsec$ (see text). }
\label{fig:MIR-radio:NGC1052}
\end{figure*}    

There is no doubt that the dust around AGN plays an important role in obscuring the central engine and eventually fueling the accretion disk \citep[][and references therein]{Netzer15}. Therefore, the distribution and composition of the dust in AGN and how they might evolve at different stages of the AGN evolution is of great importance \citep{Elitzur06,Elitzur14,Elitzur16}. This is why the torus properties in AGN have been largely discussed in the literature, mainly through unresolved SED fitting \citep[e.g.][]{Ramos-Almeida09,Ramos-Almeida11,Alonso-Herrero11,Gonzalez-Martin15, Martinez-Paredes15, Fuller16, Gonzalez-Martin17, Martinez-Paredes17}, and also with mid-IR interferometric measurements \citep{Tristram09,Tristram11,Tristram14}, since its size ($\rm{<10 pc}$) cannot be resolved with arcsecond resolution images. The only facility able to achieve sub-arcsecond angular resolution at sub-mm, and therefore to actually image (at least some) of the AGN tori is ALMA. This sub-mm interferometer provides a unique opportunity to directly measure the size of the torus. However, this wavelength band could be contaminated by the contribution of synchrotron emission by the jet and/or dust heated by star-formation \citep[][]{A-AH14, Esquej14, Esparza17, Martinez-Paredes18}. The latter might be decontaminated thanks to the super spatial resolution of ALMA data. In this work we use two main assumptions to compute the percentage of torus contribution at sub-mm wavelengths: (1) the continuum of dust at mid-IR wavelengths follows a well reported relation with the X-ray luminosity \citep{Lutz04,Gandhi09,Asmus15} and (2) the 5\,GHz luminosity is also linked to the X-ray luminosity, known the BH mass of the system \citep[known as the radio fundamental plane][]{Bonchi13}. We used the dusty torus models of \citet{Nenkova08B} for mid-IR wavelengths and a power-law distribution for cm wavelengths to extrapolate these predictions to the sub-mm wavelengths. In this way we showed that the chances to detect the torus are low. These chances slightly increase when using the lowest ALMA wavelengths (or the highest ALMA frequencies, see Fig. \ref{fig:SEDcontributors}) and high accretion sources (mainly low BH mass and high luminous sources, see Fig. \ref{fig:Jetcontributors}). Among all of these assumptions the most risky one is the radio fundamental plane. It actually traces radio-core emission however, compact new radio jet component (therefore dominating at higher radio frequency) could contribute to the radio band and, in particular, substantially to the sub-mm band. Indeed, in the previous section we show that the fundamental plane is a rough estimator of the sub-mm contribution of the jet (see Fig. \ref{fig:Radiofitting}). However, it cannot describe a possible complex behavior in the radio band, as clearly seen at least in NGC\,1052 (and perhaps in NGC\,1068, see below). Furthermore, in the particular case of NGC\,3516, the fundamental plane over-predicts the jet contribution compared to the detailed analysis of the SED. 

\begin{figure*}[!t]
\centering
\includegraphics[width=2.\columnwidth]{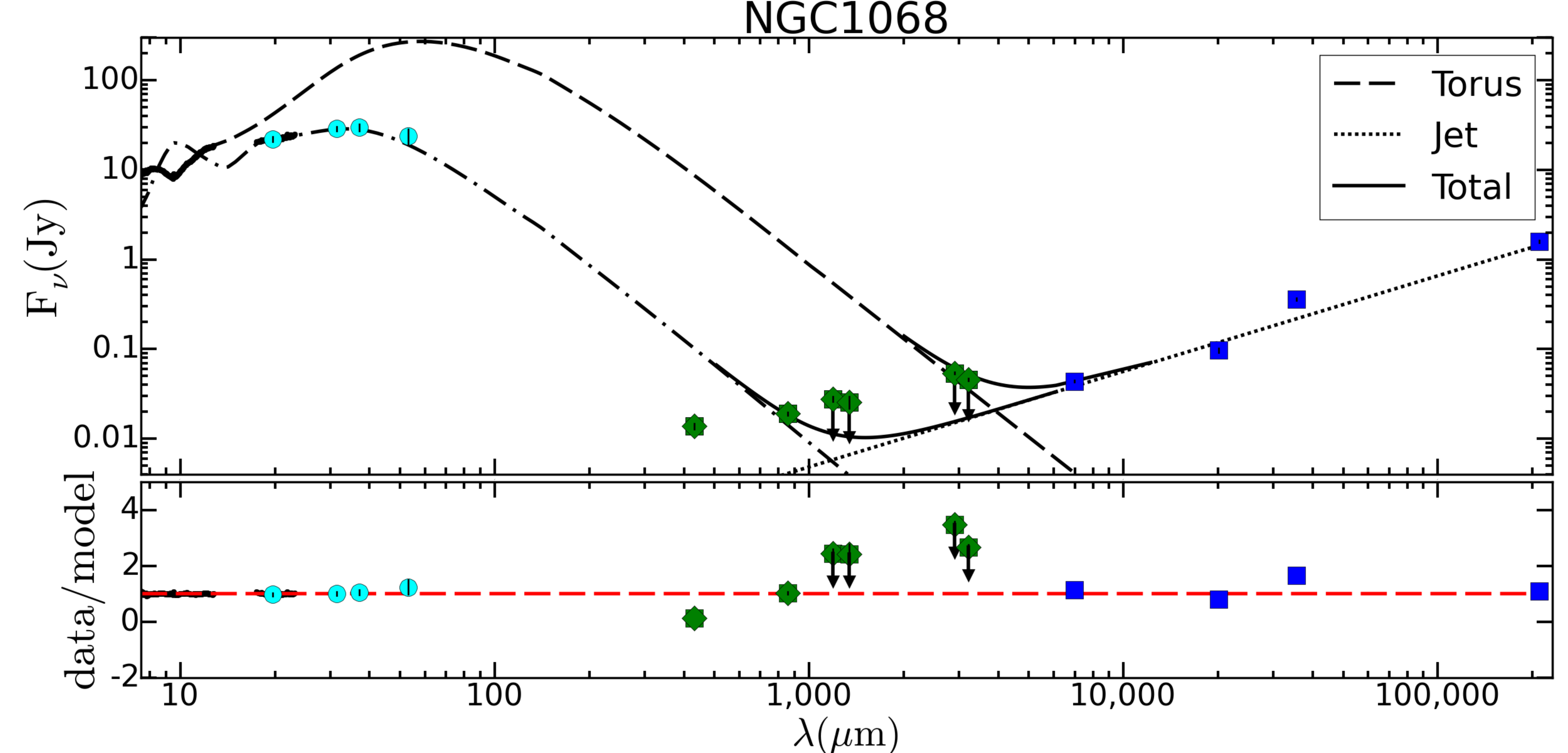}
\caption{SED fit for NGC\,1068 to the mid-IR spectra (black spectrum) with the torus model (long-dashed and dot-dashed lines show the fit to the N- and Q-band respectively, see text) and to the cm wavelengths (blue points) with synchrotron emission (dotted line). Cyan circles are the SOFIA data points. Green stars are sub-mm and far-infrared data. Arrows indicate plausible upper limits when the spatial resolution is worse than 0.3$\arcsec$ (see text). Note that residuals for the sub-mm data are computed using the Q-band mid-IR fit to the torus together with the VLA fit to the jet.}
\label{fig:MIR-radio:NGC1068}
\end{figure*}    

\begin{figure*}[!t]
\centering
\includegraphics[width=2.\columnwidth]{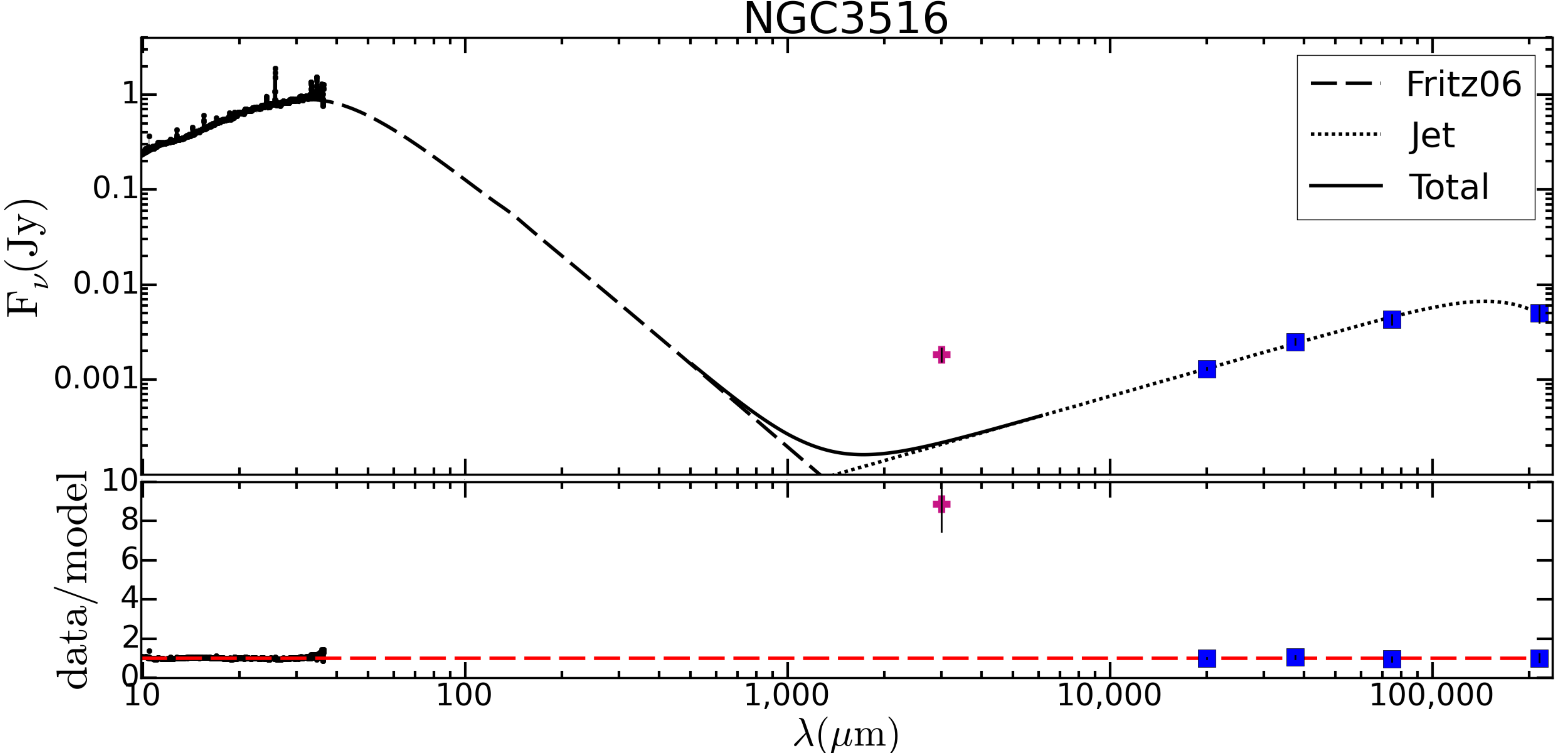}
\caption{SED fit for NGC\,3516 to the mid-IR spectra (black spectrum) with the torus model (long-dashed line) and to the cm wavelengths (blue points) with synchrotron emission (dotted line). Pink cross is sub-mm data. }
\label{fig:MIR-radio:NGC3516}
\end{figure*}

\begin{figure*}[!t]
\centering
\includegraphics[width=2.\columnwidth]{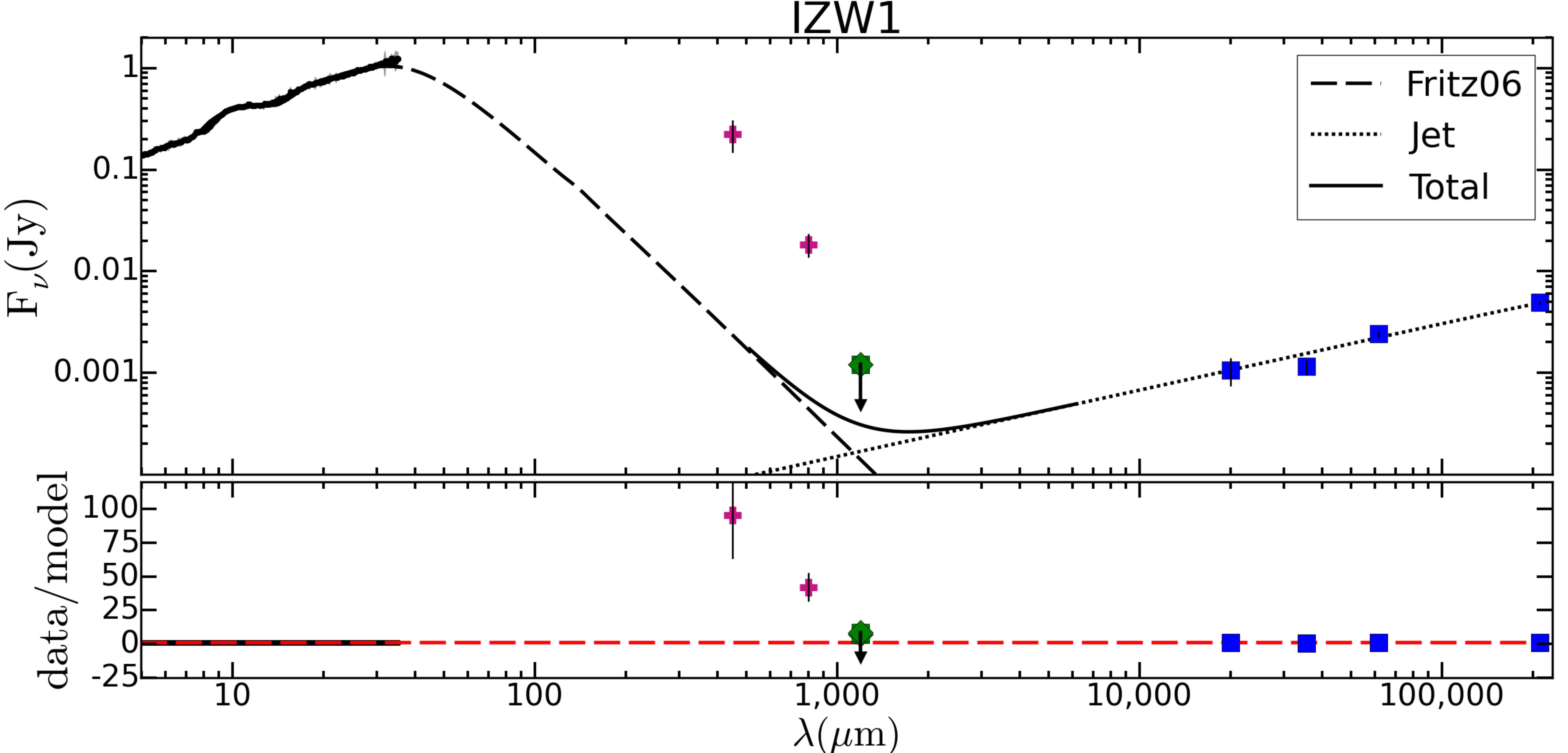}
\caption{SED fit for I\,Zw1 to the mid-IR spectra (black spectrum) with the torus model (long-dashed line) and to the cm wavelengths (blue points) with synchrotron emission (dotted line). Green star and pink crosses are sub-mm and far-infrared data, respectively. Arrows indicate plausible upper limits when the spatial resolution is worse than 0.3$\arcsec$  (see text). }
\label{fig:MIR-radio:IZW1}
\end{figure*}

In order to make a clear picture on the contributors to sub-mm wavelengths, we show the full SED fitting when combining together the mid-IR and the radio data for NGC\,1052, NGC\,1068, NGC\,3516, and IZw\,1 in Figs.\,\ref{fig:MIR-radio:NGC1052}, \ref{fig:MIR-radio:NGC1068}, \ref{fig:MIR-radio:NGC3516}, and \ref{fig:MIR-radio:IZW1}, respectively. We kept the same symbols and colors to describe the mid-IR  and radio data as those used during the separated fitting procedure. The ALMA data with angular resolution larger than 0.3$\arcsec$ are marked with a black arrow. These data points, together with JCMT and CARMA data, could be overestimating the flux density at that frequency/wavelength due to external contributors, because of the relatively large beam size. Both fits have been extrapolated in the sub-mm window and the sum of the two components are shown as black straight lines.

For the LLAGN NGC\,1052 the radio jet contribution expands towards the sub-mm window while the torus contribution falls down very quickly at $\sim$100$\mu$m (Fig.\,\ref{fig:MIR-radio:NGC1052}). This is consistent with our theoretical predictions which show that the low accreting AGN (with high BH mass and low X-ray luminosity) tend to be highly dominated by the jet at sub-mm wavelengths (see Section \ref{sec:jetcontrib}). It is also in agreement with previous analysis of ALMA data for the LLAGN NGC\,1097 \citep{Izumi17} and NGC\,1377 \citep{Aalto17}, where they report strong contributions of a variable component associated to the jet.

Indeed, the jet of NGC\,1052 dominates the vast majority of the emission in the $\rm{5- 10^6\mu m}$ range, with the overall peak of the emission at radio frequencies. This is not the case for the rest of the three sources, where the overall peak of the emission is below 50$\rm{\mu m}$. In any of these cases, neither the torus or the jet can explain solely or as combined effect the sub-mm emission with a ratio between data and model of $\rm{\sim}$3, 8 and even 100 for NGC\,1068, NGC\,3516, and I\,Zw1, respectively. 

Very complex is the situation for NGC\,1068. We could not fit the N- and Q-band mid-IR spectra to a single model. On the contrary, we needed to allow three parameters ($\gamma$, $\beta$ and $\tau$ see Tab.\ \ref{tab:MIRparam}) \footnote{Although beta was set as free parameter, both bands gave the same result.} to behave differently while the rest of the parameters are set to the same value. In this way we obtained two extrapolations of the torus component to the sub-mm wavelengths. We highlight to the reader that the fit reported here is far from realistic but is a clear indication of complex configuration of dust which cannot be fitted to a simple torus model (see Section \ref{sec:mirfit}), which is well documented in the literature \citep[see ][and references therein]{Lopez-Rodriguez16}. 

\begin{figure}[h!]
\begin{center}
\includegraphics[width=1.\columnwidth]{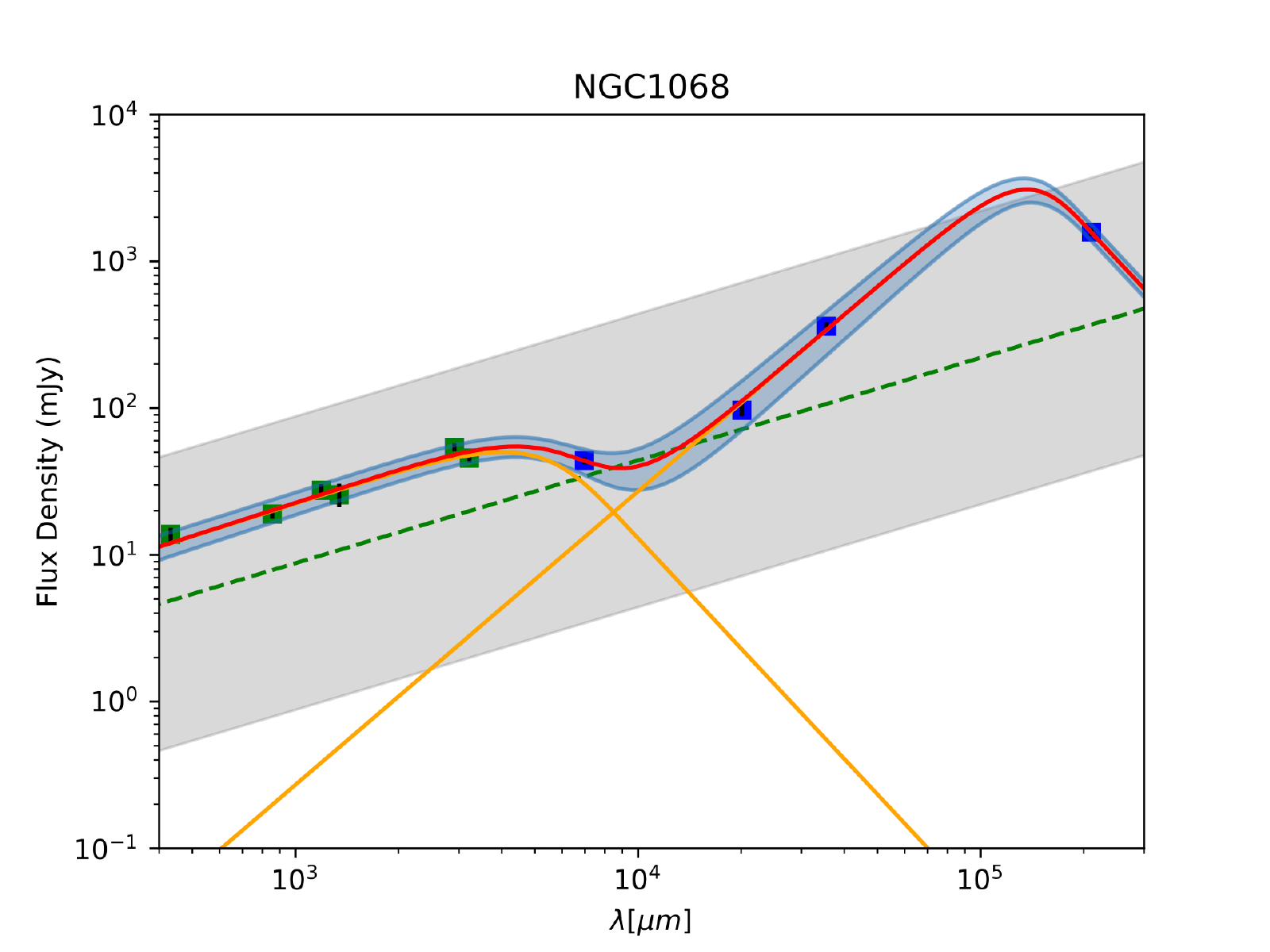}
\caption{Radio cm and sub-mm fit for NGC\,1068. Best fit and its error are shown with red straight line and transparent blue band. Orange straight lines are the two single synchrotron components, both with free spectral index values. Blue data are cm data points, green data are ALMA data points. Dashed green line and gray band are the fundamental plane prediction together with its uncertainty.}
\label{NGC1068cmsubmmdatafit}
\end{center}
\end{figure}

Note that the SOFIA data reported by \citet{Lopez-Rodriguez18} at the $\rm{20-50\mu m}$ range show the peak of the mid-IR emission in $\rm{30-40\mu m}$ range, consistent with the Q-band rather than with the N-band spectral fit. After SED fitting to mid-IR and ALMA data at $\rm{\sim 450 \mu m}$ (i.e. 666 GHz), they reported a torus outer radius of $\rm{\sim 5}$ pc. This outer radius is larger than the one reported here ($\rm{\sim 2 pc}$, see Table \ref{tab:MIRparam}). However, they argue that the use of SOFIA and ALMA data yields to larger estimate on the outer radius of the torus. This was also shown by \citet{Garcia-Burillo16}, \citet{Gallimore16}, and \citet{Imanishi18}. The question we would like to answer here is whether the sub-mm ALMA data is actually tracing the dust in the torus. Indeed the mid-IR Q-band spectral fit can easily account for the $\rm{\sim 450 \mu m}$ density flux (Fig. \ref{fig:MIR-radio:NGC1068}). Moreover, the mid-IR N-band fit could even predict the ALMA upper limits at $\rm{\sim 3000 \mu m}$. However, the entire SED at sub-mm wavelengths (with flux densities increasing with wavelength) is tough to explain with any complex scenario of dust. When considering NGC\,1068 radio cm and sub-mm data points altogether for the radio fitting, the result is a combination of two synchrotron components with two different spectral indexes representing the optically thin parts (see Fig. \ref{NGC1068cmsubmmdatafit}). The first synchrotron component peaks at 2.24 ($\pm$ 0.09) [GHz] with flux density 3088 ($\pm$ 301) [mJy] and has a spectral index value of $\alpha_{th1}$ $\sim$ --2.0 $\pm$ 0.1 ; the second component peaks at 60 ($\pm$ 2) [GHz] with a flux density of  50 ($\pm$ 2) [mJy] and a spectral index value of  $\alpha_{th2}$ $\sim$ --0.75 $\pm$ 0.03. The $\chi^{2}/dof$ value of the fit is 0.8. The former value represents an "aged" distribution of synchrotron electrons while the last is flatter. This increases the jet contribution in the sub-mm window placing further the possibility to detect the emission from the dusty torus at this wavelength band.

Also puzzling is the case of NGC\,3516. The sub-mm data point (CARMA data) does not lie on its predicted value if it is the sum of the jet and the torus contributors (large angular resolution). Thus, either the parameters of the torus obtained from the mid-IR wavelengths are wrong or there is another component contributing to the sub-mm wavelength. This extra component could be (1) dust heated by star-formation (considering the lower-spatial resolution of the sub-mm data point) or (2) a younger knot component of the jet as seen in NGC\,1052 or NGC\,1068.

In the case of I\,Zw1 the sub-mm data points are well above the prediction, most likely because of the large angular resolution. This object is clearly extended at MIR on scales of few hundred pc \citep{Martinez-Paredes17}, but it is still not clear if the extended emission is related with the possible presence of a circumnuclear ring ($\sim1.7$ kpc of diameter) detected by \citet{Schinnerer98} using $^{12}$CO(2-1) and $^{13}$CO(1-0) molecular lines. A fraction of this emission could be also be attributed to a silicate-emitting dust component that extent until the inner part of the narrow line region \citep{Scharwachter2008}. On the other hand, \citet{Hughes1993} argued that a fraction of the emission at sub-mm wavelength could be attributed to dust heated by an extended ($>1$ kpc) starburst region. Therefore, the JCMT data (i.e. the magenta crosses) could be simply explained by an additional dust contributor from star-formation, peaking around $\rm{\sim 100-200\mu m}$. Uncertain is the ALMA data at 1000$\mu$m. Although this point could be overestimating the flux density at this wavelength, it lies where the contribution of the flux density changes from the torus to the jet contributor. Dust heat by star-formation (as an extrapolation of the JCMT data) could also explain the ALMA detected flux. However, even with all the caveats in mind, this is the most promising case of torus being the dominant contributor to sub-mm wavelength because at least a large fraction of emission is expected to come from  the dusty torus. This result is consistent with our theoretical predictions because IZw\,1 is the highest accreting source among our sample (with relatively low BH mass and high X-ray luminosity, see Table \ref{tab:ObsParam}). Finally, the estimated outer radius of the torus for this source is $\rm{\sim 8}$\,pc, which at the distance of I\,Zw1 (240.7 Mpc) implies that the torus has an angular size on the sky of $\rm{\sim 0.007\arcsec}$. The ALMA image in fact shows a point-like morphology with a beam size of $\rm{\sim}$0.6$\arcsec$ \citep{Imanishi16a}. Thus, even if the torus is dominating in this high accreting sources, it might be difficult to resolve it with current interferometric observations.  
 
As a final caveat, we would like to stress that our findings on the SED fitting are biased toward the use of the library of SED models. Although we tested up to the five most used dusty models (see Section \ref{sec:mirfit}), we decided to use in all the cases the smooth torus model described by \citet{Fritz06}. The rest of the models need an extra stellar component to account for the short mid-IR wavelengths. Although, we do not expect large discrepancies using any of the other models to extrapolate the torus to the sub-mm wavelengths, all of them are (perhaps not very realistic) simplifications of the configuration of dust in AGN. This is clearly shown in the case of NGC\,1068 where a complex fit was needed to account for the spectral fit of both N- and Q-bands together. Therefore, a much more complex scenario could yield to a different contribution of the dusty torus at sub-mm wavelengths.
 Indeed, as already mentioned in Sect. \ref{sec:toruscontrib}, we also considered  the contributions from the hot graphite dust and the dusty NLR clouds \citep{Mor2009,Mor2012}.  The contribution by the former is insignificant  when extrapolated to the sub-mm window and the contribution of the last, although it could be as large as 40\% of the total emission at 24 $\mu$m, is not enough to describe the  behavior in the sub-mm band.
 
The message to take away is that unfortunately there are very few chances to detect the torus over the jet at sub-mm wavelengths. This has been claimed to be the case for NGC\,1068 \citep{Garcia-Burillo16}. They claimed the detection of the dusty torus based on a fitting of the \textit{u-v} visibilities within a radius of 0.2$\arcsec$. From this result the torus would be within 0.1$\arcsec$. However, looking at the resolution of their continuum image, the source is only 2-3 beam resolved, therefore it is marginally resolved. Moreover, although they do mention that 18\% of the total flux within that region could come from other mechanisms different from thermal dust, from our theoretical calculation we do estimate that the contribution from the synchrotron emission from the radio jet could be as high as $\sim$94\% at this high frequency. Indeed, from our SED analysis, this observations might be contaminated by the jet components. The most promising case among our targets is I\,Zw1. However, even in this case the chances to spatially resolve it are low. 
We also explored the SED of the type Seyfert 2, the Circinus galaxy. However, this is a complex source, as shown in the recent work by \cite{Izumi2018}, with a large ISM contamination in the mid-IR Spitzer spectrum. Therefore, due to the large contamination of warm dust from star formation, we could not extrapolate this contribution to the submm range. Moreover, \cite{Elmouttie1998} revealed a very flat radio source at the position of the nuclear region at frequency larger that 8 GHz. Therefore, also in this case, it is plausible that there is still synchrotron emission contamination in the sub-mm band, due to high frequency radio components from the radio jet.
We strongly suggest to perform SED analysis, including radio data, with particular attention to the angular resolution used, before drawing any conclusion on the detection of a torus at sub-mm wavelengths. Another possibility already exploited by other authors \citep[e.g.][]{Izumi17} is to study the variability pattern of the sub-mm wavelengths because we do not expect the torus to vary in a year-period basis while the jet would show such variations.  





\section{Summary}\label{sec:summary}

This work aims to study the detectability of the torus in the sub-mm window. Specifically, we wanted to analyze this aspect considering the ALMA capabilities. Indeed, this sub-mm interferometer is the only facility able to achieve sub-arcsec angular resolution and very good sensitivity (tens of $\mu$Jy) to actually image the AGN tori. We hack this issue using two approaches: the first is a theoretical approach and the second an observational one.

For the former, we considered the most used SED models for dusty AGN emission as the Clumpy models \citep{Nenkova08A,Nenkova08B} together with known scaling relations between the accretion disk luminosity and the torus continuum \citep[][]{Gandhi09,Netzer15,Asmus15}, and the accretion disk luminosity, jet contribution and SMBH mass \citep[][]{Bonchi13,Saikia15}. We extrapolated the scaling relations to the sub-mm wavelengths and we estimated in which conditions the torus could prevail on the radio jet. The main result is that we are more likely to detect bigger and denser dusty tori at the highest ALMA frequency (666\,GHz/450\,$\mu$m).  A 1 mJy torus (characteristics of bright AGN, with flux density S$\rm{\sim}$10\,Jy) can be detected with high detection limit, observing:  1h at 353\,GHz/850\,$\rm{\mu m}$ and 10h at 666\,GHz/\,450$\rm{\mu m}$, being aware of the possible contamination by the radio jet.

For the observational approach, we used four prototypical AGN: NGC\,1052 (LLAGN), NGC\,1068 (Type-2 Seyfert), NGC\,3516 (Type-1.5 Seyfert) and I\,Zw1 (QSO). These targets have been selected to have available radio data (to trace the jet contribution), mid-IR data (to trace the torus contribution), and continuum sub-millimeter observations. After performing individual mid-IR \citep[using the smooth dusty model described by ][]{Fritz06} and cm-radio (using a power law and a single, or a composition of, synchrotron component/s) fit to the spectra, we combined them together to compare the extrapolation of both torus and jet contributors at sub-mm wavelengths with data at those wavelengths. The radio jet of NGC\,1052 dominates over the torus in the sub-mm range. In all the other cases, neither the torus or the jet can explain the sub-mm emission. This result is consistent with the theoretical predictions for which: low accretion AGN are highly dominated by the jet, while high accretion sources (like I\,Zw 1) might be good candidates to detect emission from the torus at sub-mm wavelengths.

We therefore suggest to perform SED analysis including radio data in order to isolate the torus contribution from jet or dust heat by the AGN.

\section*{Acknowledgements}
{\footnotesize This paper is mainly funded by the DGAPA PAPIIT project IA103118.
This paper makes use of the following ALMA data: \\2012.1.00034.S, 2013.1.01225.S, 2013.1.00055.S, 2013.1.01151.S, 2013.1.00221.S, 2015.1.00989.S, 2015.1.00591.S, 2016.1.00232.S. ALMA is a partnership of ESO (representing its member states), NSF (USA) and NINS (Japan), together with NRC (Canada), MOST and ASIAA (Taiwan), and KASI (Republic of Korea), in cooperation with the Republic of Chile. The Joint ALMA Observatory is operated by ESO, AUI/NRAO and NAOJ. \\
Support for CARMA construction was derived from the states of California,
Illinois, and Maryland; the James S. McDonnell Foundation; the Gordon and
Betty Moore Foundation; the Kenneth T. and Eileen L. Norris Foundation; the
University of Chicago; the Associates of the California Institute of
Technology; and the National Science Foundation. Ongoing CARMA
development and operations are supported by the National Science Foundation
under a cooperative agreement, and by the CARMA partner universities.\\
The James Clerk Maxwell Telescope is operated by the East Asian Observatory on behalf of The National Astronomical Observatory of Japan; Academia Sinica Institute of Astronomy and Astrophysics; the Korea Astronomy and Space Science Institute; the Operation, Maintenance and Upgrading Fund for Astronomical Telescopes and Facility Instruments, budgeted from the Ministry of Finance (MOF) of China and administrated by the Chinese Academy of Sciences (CAS), as well as the National Key R\&D Program of China (No. 2017YFA0402700). Additional funding support is provided by the Science and Technology Facilities Council of the United Kingdom and participating universities in the United Kingdom and Canada.\\
This research has made use of the NASA/IPAC Extragalactic Database (NED),
which is operated by the Jet Propulsion Laboratory, California Institute of Technology,
under contract with the National Aeronautics and Space Administration.\\
A.P. acknowledges support by UNAM-DGAPA and to C\'atedra CONACyT}. D.E.A. and N.O.C. would like to acknowledge CONACYT scholarships No. 592884 and No. 897887, respectively. MM-P acknowledges support by UNAM-DGAPA and KASI PostDoc fellowship. 


\clearpage
\appendix
\section{Radio cm and sub-mm information}
\label{RadioInfoAppendix}
Here we report the radio cm and sub-mm information such as: the used radio cm and sub-mm telescope band and frequency, the archival project ID,  literature references together with technical information (in Tab.s \ref{radioinfo} and \ref{sub-mminfo}). Moreover, the radio cm and sub-mm flux densities are reported in Tab. \ref{FluxDensityAGN}
\begin{table*}[b]
\caption{JVLA A-configuration and literature references information.}
\label{radioinfo}
\begin{center}
\scriptsize 
\begin{tabular}{ ccccccc} 
\hline\hline
Source Name & Band & $\nu$ [GHz] & Project ID & Clean beam [$\arcsec$] & PA [deg] &  Reference \\
\hline
\multirow{9}{4em}{NGC\,1052}& --&	 0.08 & -- &	-- & -- & \cite{Slee1977}$^{(1)}$	\\
							& L	&	1.4		&	16B-289	&	2.8$\times$1.0 & --47 & --	\\	
							& C	&	4.4		& -- &	-- & -- &	\cite{Perley1982}$^{(2)}$\\	
							& C	&	6.2		&	16B-343	&	0.4$\times$0.3& --30 & --	\\	
							& X	&	8.4		& AW278		&0.3$\times$0.2 &	--24	& -- \\
							& Ku&	14.9	&	AF339	&	0.1$\times$0.09 & --30 & --	\\	
							& K	&	22.1	&	BC0066	&	0.09$\times$0.07 & 4& --	\\	
							& -- &	31.4	& -- &	-- & -- &	\cite{Geldzahler1981}$^{(1)}$\\	
\hline
\multirow{4}{4em}{NGC\,1068}& L	&	1.4		&	AU079	&	1.5$\times$1.3 & --24 & --	\\
							& X	&	8.4		&	AC467	&	0.2$\times$0.2 & 4 & --	\\
							& Ku&	14.9	&	AC467	&	0.1$\times$0.1 & --75 & --	\\ 
							& Q	&	43.0	& AC0565 &	-- & -- &	 \cite{Cotton2008}$^{(2)}$\\     
\hline
\multirow{4}{4em}{NGC\,3516} 	& L	& 1.4 & -- &	-- & -- & \cite{Ulvestad1984}$^{(2)}$\\
                    		& C	& 4.8 	& -- &	-- & -- & \cite{Ulvestad1984}$^{(2)}$	\\
                    		& X	& 8.4 	& AF360 &	-- & -- & \cite{Mundell2009}$^{(2)}$	\\
                    		& Ku & 15.0	& AW0126 &	-- & -- & \cite{Ulvestad1989}$^{(3)}$	\\
\hline\multirow{5}{4em}{I Zw 1} & L	& 1.4 & AK406 &	1.3$\times$1.2 & 3 & --	\\
                           & C	& 4.8	& AK406 &	0.4$\times$0.3 & --2 & --	\\
                           & X	& 8.4	& AB0670 &	0.2$\times$0.2 & --9 & --	\\
                           & Ku & 15.0	& AA0048 &	-- & -- & \cite{BarvainisAntonucci1989}$^{(4)}$	\\

\hline\hline

\end{tabular}

\end{center}
        \begin{flushleft}
Note: 
\newline
(1): two low angular resolution data points. These values are needed at the time of fitting the radio SED: the 80 MHz data point (from radioheliograph Culgoora-3) helps to search for the low/mid radio frequency synchrotron component. The 31 GHz observation (from single dish NRAO 11-m telescope at Kitt Peak) helps to find a good turn over point at high radio frequency.
(2): data obtained using JVLA at A configuration.
(3): observation using JVLA at B configuration, angular resolution at this frequency 0.4$\arcsec$.
(4): this flux density has been obtained from VLA observation at C configuration for which the angular resolution is 1.4$\arcsec$. In this work, the continuum flux densities at L band (6.22 + 0.35) and C band (2.21 + 0.11) are also reported. These values are in agreement with higher angular resolution L band (4.91 +- 0.12) and C band (2.41 +- 0.12) from the project AK406. Therefore, we can assume that the Ku band flux density obtained using C configuration well represent the core flux density at this frequency.   
\end{flushleft}
\end{table*}

\begin{table*}
\caption{Sub-mm information.}
\label{sub-mminfo}
\begin{center}
\scriptsize 
\begin{tabular}{ cccccc} 

\hline\hline
Source Name & $\nu$ [GHz] & Project ID & Clean beam [$\arcsec$]& PA [deg] &  Reference \\
                           
\hline
\multirow{6}{4em}{NGC\,1052}&	108 	& 2015.1.00591.S  &	0.7$\times$0.6 & --65 & --	\\
							&	113		& 2015.1.00989.S	&	1.4$\times$1.3 & --20 & --	\\	
							&	214		& 2013.1.01225.S	&	0.3$\times$0.2 & 54 &	--\\	
							&	230		& 2013.1.01225.S	&	0.3$\times$0.2& 54 & --	\\	
							&	340 	& 2013.1.01225.S	&	0.20$\times$0.15 & 57	& -- \\
							&	350		& 2013.1.01225.S	&	0.20$\times$0.15 & 57 & --	\\	
\hline
\multirow{5}{4em}{NGC\,1068}&	93	&	2013.1.00055.S	&	0.7$\times$0.5 & 74 & --	\\
							&	103	&	2013.1.01151.S	&	4.6$\times$2.1 & 67 & --		\\ 
							&	223	& 	2016.1.00232.S	&	0.3$\times$0.3& 80 &	 --\\     
							&	251	&	2013.1.00221.S	&	0.5$\times$0.5 & 60 & --	\\
							&	350	& 	2016.1.00232.S 	&	0.2$\times$0.2 & 77 &	 --\\     
							&	694	& 	2013.1.00055.S 	&	0.07$\times$0.05& 60 &	\cite{Garcia-Burillo16}\\     
\hline
\multirow{1}{4em}{NGC\,3516} & 100 & -- &	-- & -- & \cite{Behar2018}$^{(1)}$\\
\hline
\multirow{3}{4em}{I Zw 1}  &	252 & 2012.1.00034.S &	0.7$\times$0.6  & 60 & \cite{Imanishi2016b}\\
                           &374	& -- &	-- & -- & \cite{Hughes1993}$^{(2)}$ \\
                           &666	& -- &	-- & -- & \cite{Hughes1993}$^{(2)}$ \\
                           
\hline\hline

\end{tabular}

\end{center}
        \begin{flushleft}
Note: 
\newline
 (1) Data using CARMA telescope at C-array configuration, providing an angular resolution of $\sim$1$\arcsec$. 
 (2) Data using JCMT telescope, providing an angular resolution at 374 GHz and at 666 GHz of 13.5$\arcsec$ and 9$\arcsec$ respectively.
\end{flushleft}
\end{table*}

\begin{table}[h!]
\caption{Radio and sub-mm flux densities of the targets.}
\label{FluxDensityAGN}

\begin{center}
\scriptsize 
\begin{tabular}{ ccccc} 

\hline\hline
Source Name & $\nu$ & S [mJy]& errS & Note\\

\hline
multirow{14}{4em}{NGC1052}	&	 0.08 &	4000	&	400 &$^{(1)}$ \\
		&	1.4		&	663.3	&	2.0&		\\	
		&	4.8		&	1130.0	&	113.0&$^{(2)}$\\	
		&	6.2		&	1295.0	&	1.2 &	\\	
		&	8.4  &	2910.0  & 	30 &		\\
		&	14.9	&	1020.0	&	100.0&	\\	
		&	22.1	&	2460.0 	&	 20.0&	\\	
		&	31.4	&	1070.0	&	120.0&$^{(3)}$\\	
		&	108.0	&	750.0	&	7.4 &	 \\	
		&	113.7	&	773.0	&	3.0 &	\\	
		&	214.1	&	495.1  	&	 3.4 &	\\	
		&	229.4	&	456.6  	&	 4.0 &	\\	
		&	340.8	&	299.3  	&	 1.3 &	\\	
		&	352.8	&	283.1  	&	 1.1 &	\\	
\hline
\multirow{9}{4em}{NGC1068}	&	1.4	&	1585.0	&	51.0& \\
			&	8.4		&	361.0	&	12.0& \\
			&	14.9	&	97.0	&	6.0 & \\ 
			&	43.0	&	44.0	&	1.0 &$^{(4)}$ \\     
			&	93.1	&	46.0	&	2.0 &\\ 
			&	103.0	&	54.0	&	2.0& \\ 
			&	223.6	&	26.0	&	4.0& \\ 
			&	252.0	&	28.0	&	2.0& \\ 
			&	351.0	&	19.2	&	1.0&\\
			&	694.0	&	13.8	&	1.0&$^{(*)}$\\
\hline
\multirow{5}{4em}{NGC3516} & 1.4 & 5.0 & 1.0& $^{(5)}$	\\
                    & 4.0 	&	4.30	&	0.50 &$^{(5)}$	\\
                    & 8.0 	&	2.50	&	0.20 &$^{(6)}$	\\
                    & 15.0	&	1.30	&	0.03 &$^{(7)}$	\\
					& 100	& 1.84		& 0.29	&$^{(8)}$\\ 
\hline
\multirow{7}{4em}{I\,Zw1} & 1.4 & 4.91 &	0.12 &	\\
                            & 4.8	& 2.41 & 0.12&	\\
                            & 8.4	& 1.15 & 0.17 &	\\
                            & 15.0	& 1.06 & 0.31&$^{(9)}$	\\
                            & 252.0	& 1.20 & 0.06&$^{(10)}$	\\
                            & 374	& 18.4	&	4.4 &$^{(11)}$ \\
							& 666	& 225	&	75	& $^{(11)}$ \\
\hline\hline

\end{tabular}
\end{center}
        \begin{flushleft}

Note: 
(1) from \cite{Slee1977};
(2) from \cite{Perley1982};
(3) from \cite{Geldzahler1981};  (4) from \cite{Cotton2008}; (5) from \cite{Ulvestad1984}; (6) from \cite{Mundell2009}; (7) from \cite{Ulvestad1989};(8) from \cite{Behar2018}; (9) from \cite{BarvainisAntonucci1989}; (10) from \cite{Imanishi2016b}; (11) from \cite{Hughes1993};  (*) flux density from \cite{Garcia-Burillo16}: at the resolution 0.07$^{\arcsec}$ $\times$ 0.05$^{\arcsec}$ this value correspond to the flux density of the claimed NGC\,1068 core S1 component \cite{Gallimore2004}
\end{flushleft}
\end{table}
\clearpage


\end{document}

